\documentclass[journal]{IEEEtran}
\usepackage[dvips]{graphicx}
\usepackage{amsmath}
\usepackage{amsthm}
\usepackage[flushleft]{threeparttable}
\usepackage{array}
\usepackage{booktabs}
\newcommand{\PreserveBackslash}[1]{\let\temp=\\#1\let\\=\temp}
\newcolumntype{C}[1]{>{\PreserveBackslash\centering}p{#1}}
\newcolumntype{R}[1]{>{\PreserveBackslash\raggedleft}p{#1}}
\newcolumntype{L}[1]{>{\PreserveBackslash\raggedright}p{#1}}

\usepackage{bm}
\usepackage{graphicx,subfigure}
\usepackage{algorithm}
\usepackage{cite}
\usepackage{amsmath}
\usepackage{amssymb}
\usepackage{psfrag}
\usepackage{algpseudocode}
\usepackage{empheq}
\usepackage{latexsym, amsmath, subfigure, color, amsfonts, amssymb,graphicx}
\usepackage{wrapfig} % Allows wrapping text around tables and figures
\newtheorem{Theorem}{Theorem}
\newtheorem{Lemma}{Lemma}
\newtheorem{Definition}{Definition}
\newtheorem{Corollary}{Corollary}

\newtheorem{Remark}{Remark}
\IEEEoverridecommandlockouts
\usepackage{tabularx}
\DeclareMathOperator*{\argmin}{\arg\!\min}

\usepackage[T1]{fontenc}
\usepackage{etoolbox}

\makeatletter
\def\hlinewd#1{%
\noalign{\ifnum0=`}\fi\hrule \@height #1 %
\futurelet\reserved@a\@xhline}
\patchcmd{\maketitle}{\@fnsymbol}{\@alph}{}{}  % Footnote numbers from symbols to small letters
\makeatother

\title{Coded Caching and Content Delivery with Heterogeneous Distortion Requirements}
\author{
Qianqian Yang,~\IEEEmembership{Student Member,~IEEE}, and\thanks{The authors are with Imperial College London, London SW7 2AZ, U.K. (e-mail: q.yang14@imperial.ac.uk; d.gunduz@imperial.ac.uk).

This paper was presented in part at the 2016 IEEE International Symposium on Information Theory in Barcelona, Spain \cite{QianqianlossyCaching}. 

This work has been partially funded by the European Research Council (ERC) under the European Union's Horizon 2020 research and innovation programme through Starting Grant BEACON (agreement No. 677854).

}
  \and
  Deniz G\"und\"uz,~\IEEEmembership{Senior Member,~IEEE}
}
\date{}
\begin{document}
\pagestyle{empty}
\maketitle
\begin{abstract}
Cache-aided coded content delivery is studied for devices with diverse quality-of-service (QoS) requirements, specified by a different average distortion target. The network consists of a server holding a database of independent contents, and users equipped with local caches of different capacities. User caches are filled by the server during a low traffic period without the knowledge of particular user demands. As opposed to the current literature, which assumes that the users request files in their entirety, it is assumed that the users in the system have distinct distortion requirements; and therefore, each user requests a single file from the database to be served at a different distortion level. Our goal in this work is to characterize the minimum \textit{delivery rate} the server needs to transmit over an error-free shared link to satisfy all possible demand combinations at the requested distortion levels, considering both \textit{centralized} and \textit{decentralized} cache placement. 
   
For centralized cache placement, the optimal delivery rate is characterized for the two-file two-user scenario for any pair of target distortion requirements, when the underlying source distribution is successively refinable. For the two-user scenario with more than two successively refinable files, the optimal scheme is characterized when the cache capacities of the users are the same and the number of files is a multiple of $3$. For the general source distribution, not necessarily successively refinable, and with arbitrary number of users and files, a \textit{layered} caching and delivery scheme is proposed, assuming that scalable source coding is employed at the server. This allows dividing the problem into two subproblems: the lossless caching of each layer with heterogeneous cache sizes, and cache allocation among layers. A delivery rate minimization problem is formulated and solved numerically for each layer; while two different schemes are proposed to allocate user caches among layers, namely, \textit{proportional cache allocation (PCA)} and \textit{ordered cache allocation (OCA)}. A decentralized lossy coded caching scheme is also proposed, and its delivery rate performance is studied. Simulation results validate the effectiveness of the proposed schemes in both settings.
\end{abstract}
% ==================================================
\section{Introduction}
Consumer demand for mobile video services is growing at an unprecedented rate. This trend is expected to continue in the coming years driven by the proliferation of mobile devices with high quality display capabilities, and the explosion of high data rate multimedia contents available online. The majority of video traffic today is generated through ``video surfing'', i.e., streaming of video files stored in large databases, e.g., YouTube, Hulu, Dailymotion, BBC iPlayer, etc. Such video traffic is dominated by a relatively small number of viral contents that remain popular over a certain period of time, and downloaded repeatedly by many users, sometimes connected through the same access point, creating a huge amount of repetitive traffic both on the core and the radio access networks. 

Caching has long been used in the Internet to reduce traffic, as well as latency (see \cite{fan2000summary}, and references therein). More recently, research on content caching has regained popularity, targeting mainly wireless networks (see \cite{golrezaei2012femtocaching,AlmerothCacing, Gregori2015multi}, and references therein). While content caching at the evolved packet core, or at the radio access network, can reduce traffic and latency on the backhaul links~\cite{golrezaei2012femtocaching, wang2014cache}, caching contents directly at user devices can bring further benefits \cite{Gregori2015multi, maddah2014fundamental}. The latter strategy is called \textit{proactive caching} since caching at a user device requires predicting the demand, and delivering the content even before it is requested by the user. Since mobile data traffic at wireless access points exhibits a high degree of variation across time, exploiting the radio resources during low traffic periods through proactive caching will also reduce the peak traffic rates; and therefore, improve the quality-of-experience (QoE) for users, and reduce infrastructure costs for network providers. Feasibility of proactive caching in future wireless networks is further supported by the low-cost and abundance of storage space in today's mobile devices. 

In the proactive caching model considered here, users fill their caches during the off-peak traffic period, referred to as the \textit{placement phase}. Each user's cache content at the end of the \textit{placement phase} is a function of the whole database. User requests, one file per user, are revealed during the peak traffic period, and satisfied simultaneously during the \textit{delivery phase}. Conventional \textit{uncoded caching} schemes store contents, partially or fully, at each user's cache, and utilize orthogonal unicast transmissions during the delivery phase. The gain from uncoded caching for each user depends only on the local cache capacity. Recently, Maddah-Ali and Niesen introduced a novel model for cache networks \cite{maddah2014fundamental}, particularly appropriate for proactive caching in wireless networks, in which the delivery phase is carried out over a shared link, modeling the broadcast nature of wireless communications. They show that \textit{centralized coded caching}, in which coded bits, rather than plain bits of the available contents in the database, are cached and delivered, achieves significant gains compared to uncoded caching. This gain is realized by jointly optimizing the placement and the delivery phases to create multicasting opportunities even among distinct requests~\cite{maddah2014fundamental}. In contrast to the \textit{centralized setting} in~\cite{maddah2014fundamental}, where the active users are known in advance, \cite{maddah2013decentralized} considers the so-called \textit{decentralized setting}, in which, during the \textit{placement phase}, the server has no prior knowledge on the number and identity of users that will participate in the \textit{delivery phase}. It is shown that the multicasting opportunities still appear even if users randomly cache bits of files \cite{maddah2013decentralized}. 

Numerous papers followed \cite{maddah2014fundamental} and \cite{maddah2013decentralized} in order to further improve the \textit{coded caching} gain, and to apply it to various other network models. Chen et al.~\cite{chen2014fundamental} achieve the optimal delivery rate for small buffer sizes by placing coded contents into users' caches during the \textit{placement phase}. When the number of users is larger than the number of files, improved delivery rates are obtained in~\cite{mohammadqian2016large, PabloCaching}. Pedarsani et al. introduce a coded least-recently sent delivery and update rule that replaces the cache content during the delivery phase for online caching systems~\cite{pedarsani2014online}. A multi-layer caching system, in which user terminals, proxies, base stations are all equipped with cache memories, is considered in~\cite{karamchandani2014hierarchical}. A distributed caching system is investigated in~\cite{Caire2015distributing} with single-hop device-to-device communication, which shows that coded caching has the same scaling law as the spatial reuse of user caches. Delivery over a noisy broadcast channel is considered in~\cite{TimoErasureChannel,amirierasure2018Tcom, amirigaussian2018, EliawirelessBC}. Similarly, delivery of contents over an interference channel is considered in~\cite{Maddahaliinterference, joan2017icc, Simeonecacheaided, Meitxrx}, where both the transmitters and receivers have caches. 

Common to the aforementioned works and most of the other follow-up papers in the literature, is the assumption that the files in the database have fixed sizes, and each user requests one of these files in whole. However, in practice, video contents are usually downloaded at different quality levels by users, which may be due to their viewing preferences, or the display and processing capabilities of their devices. For example, a laptop may require high quality descriptions of requested files, whereas a mobile phone is satisfied with much lower resolution. In current video coding standards, diverse reconstruction capacities and demands of users is handled through scalable video coding (SVC). The H.264/MPEG-4 standard~\cite{stockhammer2011dynamic, schwarz2007overview} allows temporal (frame rate), spatial (picture size) or SNR/quality scalability. This is achieved by encoding the videos into multiple bit streams, i.e., substreams, such that the more substreams users receive, the higher the corresponding resolution is. In this work, we consider users with heterogeneous quality-of-service (QoS) requirements, that is, each user, instead of requesting a file in the database in full, may request a lower resolution copy. Accordingly, we exploit scalable compression of the files available in the database. This provides flexibility to the server not only in supporting the multiple quality levels requested by the users, but also in exploiting the different cache capacities of the users.

In particular, we consider the lossy version of the coded caching problem, such that each user has a preset average distortion requirement. This distortion target is user-dependent, and is the same for any file the user may request. In centralized caching, it is assumed that the server knows the distortion requirements of all the users in the system during the \textit{placement phase}, whereas no such knowledge is needed in decentralized caching. Given the cache capacities and the distortion requirements of the users, the objective of the server is to design the placement and the delivery phases jointly, in order to minimize the delivery rate while guaranteeing that all possible demand combinations can be satisfied at the required distortion levels. 
%For the sake of mathematical analysis, the files in the database are modeled as independent sequences of Gaussian distributed random variables, and we exploit successive refinability\cite{equitz1991successive} of Gaussian sources in our derivations. 

The main contributions of this paper can be summarised as follows:

\begin{itemize}
  \item The model studied in this paper generalizes the original proactive caching model introduced in \cite{maddah2014fundamental} in two directions: the users are equipped with cache memories of different capacities, and each user may have a different QoS requirement, which translates into a different average distortion target.
  
  \item We derive a theoretical lower bound on the delivery rate based on cut-set arguments.
  
  \item For the centralized lossy caching problem, we characterize the optimal delivery rate for the two-file two-user case when the underlying distribution of the files is successively refinable for the desired distortion measure.
  
  \item We propose a centralized coded caching scheme for the case with two users and an arbitrary number of files, which is proven to be optimal for a successively refinable distribution, when the cache capacities of the users are the same, and the number of files is a multiple of $3$.
  
  \item For the general case, we propose a coded caching algorithm based on scalable coding of the files in the database into as many layers as the number of different distortion requirements.  We then divide the problem into two subproblems: the lossless caching problem of each layer with heterogeneous cache sizes, and the cache allocation problem among different layers. The first subproblem is formulated as an optimization problem using the achievable delivery rates available in the literature \cite{maddah2014fundamental, mohammadqian2016large, chen2014fundamental}, and solved numerically. Then, two cache allocation algorithms, i.e., \textit{proportional cache allocation} and \textit{ordered cache allocation}, are proposed, and their performances are compared with each other and the theoretical lower bound through numerical simulations.
 
  \item We propose a coded caching scheme for the \textit{decentralized} lossy caching problem, and derive its delivery rate.
\end{itemize}

The most related work to this paper is~\cite{hassanzadeh2015distortion}, which exploits the decentralized coded caching scheme proposed in \cite{mingyue2015orderoptimal}, and optimizes the allocation of cache capacities to minimize the average distortion across users, constrained by the delivery rate over the shared link and the cache capacities of the users. Scalable coding of the contents is also considered in~\cite{hassanzadeh2015distortion}. Instead of scalable coding, Cacciapuoti et al. consider multiple description coding in \cite{cacciapuoti2016speeding}, where the reconstruction quality depends solely on the number of received descriptions, irrespective of the order of these descriptions. The authors introduce a channel-aware caching scheme where the reconstruction qualities across users are decided according to the cache configuration and channel states. In \cite{timo2015rate}, Timo et al. also consider lossy caching, taking into
account the correlation among the available contents, based on which the tradeoff between the compression rate, reconstruction distortion and cache capacity is characterized for a single-user scenario, as well as some special cases of the two-user scenario.

The rest of the paper is organized as follows. We present the system model and the problem formulation in Section~II. Centralized coded caching is studied in Section~III, and a theoretical lower bound as well as achievable schemes are proposed. Decentralized lossy coded caching is considered in Section~IV. Numerical results are presented in Section~V. Finally, we conclude the paper in Section VI, followed by the appendices.

\section{System Model}\label{sec1}

We consider a server holding $N$ independent source sequence, $S^n_1$, ..., $S^n_N$, which may correspond to $N$ video files. Each $S^n_i$ sequence consists of $n$ independent and identically distributed (i.i.d) source samples $(S_{i,1}, ..., S_{i,n})$, for $i=1, ..., N$. There are $K$ users in the system that may request any of the files from the server. The operation of the caching system consists of two distinct phases: \emph{placement phase} and \emph{delivery phase}. In the \emph{placement phase}, users pre-fetch bits from the server to fill their caches. In \textit{centralized caching}, active users that will participate in the \emph{delivery phase} are known in advance during the \emph{placement phase}, enabling coordination of cache placement across users. On the contrary, in \textit{decentralized caching}, users fill their caches from different servers through different access points, and equivalently, the server has no prior knowledge of the active users that will participate in a particular delivery phase. Therefore, the cache placement of each user is conducted independently.

We assume that each user is equipped with a cache of size $M_kn$ bits, $k=1, ..., K$, and denote by $Z_k$ the contents of the cache of user $k$ at the end of the \emph{placement phase}.  The \emph{delivery phase} starts after users reveal their demands, denoted by $\mathbf{d}\triangleq(d_1, ..., d_K)$, where  $d_k\in \{1, ..., N\}$ denotes the demand of user $k$.  During this phase, a single message $X^n_{(d_1, ..., d_K)}$ of size $nR$ bits is sent by the server over the shared link depending on all the users' requests and cache contents. User $k$ reconstructs its requested file by combining $Z_k$ and $X^n_{(d_1, ..., d_K)}$.

An $(n, M_1, ..., M_K, R)$ ``caching code'' consists of $K$ cache placement functions:
\begin{equation}
  f^n_{k}: \underbrace{\mathbb{R}^n \times  \cdots \times \mathbb{R}^n}\limits_{N~files} \rightarrow \{1, ..., 2^{nM_k}\}~~\mbox{for}~~k=1, ..., K,  
\end{equation}
one delivery function:
\begin{equation}
g^n: \underbrace{\mathbb{R}^n \times  ... \times \mathbb{R}^n}\limits_{N~files} \times \underbrace{\{1, ..., N\}^K}\limits_{K~requests}  \rightarrow \{1, ..., 2^{nR}\},
\end{equation}
where $Z_k^n=f_k^n(S_1^n, ..., S_N^n)$, $X^n_{(d_1, ..., d_K)}=g^n(S_1^n, ..., S_N^n, d_1, ..., d_K)$, and $K$ decoding functions:
\begin{equation}
h_k^n :  \{1, ..., N\}^K \times \{1, ..., 2^{nM_k}\} \times \{1, ..., 2^{nR}\} \rightarrow \mathbb{R}^n,
\end{equation}
where $\hat{S}_{k}^n = h_k^n(\mathbf{d}, Z_k^n, X^n)$. Note that, in this formulation the demand vector $\mathbf{d}$ is known by all the users.

The distortion between the original sequence $S^n_i$ and its reconstruction $\hat{S}^n_i$ is measured by the same distortion function $d(\cdot, \cdot)$ at all receivers: 
\begin{equation}
d(S^n_i, \hat{S}^n_i)=\frac{1}{n}\sum\limits_{j=1}^{n} d(S_{ij}, \hat{S}_{ij}),
\end{equation}
where $d: \mathbb{R}\times\mathbb{R} \rightarrow [0, \infty)$ is a per-letter distortion measure. We assume that each user has a preset distortion requirement $D_k$, $k=1, ..., K$. Without loss of generality, we assume that $D_1 \geq D_2 \geq \cdots \geq D_K$. We emphasize that the distortion tuple $\mathbf{D}\triangleq(D_1, ..., D_K)$ is known during the \emph{placement phase}, while the demand realization, $\mathbf{d}$, is unknown.
 
\begin{Definition}
A distortion tuple $\mathbf{D}$ is \textit{achievable} if there exists a sequence of caching codes $(n, M_1, ..., M_K, R)$, such that
 \begin{equation}
\limsup_{n \rightarrow \infty} \mathbb{E}\Big[d(S^n_{d_k}, \hat{S}^n_{d_k})\Big] \leq D_k,~~k=1, 2, ..., K,
\end{equation}
holds for all possible request combinations $\mathbf{d}$, where the expectation is with respect to the distribution of $S_{d_k}$.
\end{Definition}

\begin{Definition}
For a given distortion tuple $\mathbf{D}\triangleq(D_1, ..., D_K)$, the \textit{cache capacity-delivery rate tradeoff} is defined as follows:
\begin{align}\label{eq2}
R^\star(M_1, ..., M_K) \triangleq \inf\{R: \mathbf{D}~\mbox{is~achievable}\}.
\end{align}
\end{Definition}

For ease of exposure, we assume that all the files in the library follow the same distribution such that the compression rate for a fixed distortion target, that is, the rate-distortion function, is identical for all the files. We denote by $r^*_k$ the minimum compression rate that achieves an average distortion of $D_k$, which is given as follows, in a single-letter form,
\begin{equation}
r^*_k\triangleq R(D_k)= \min\limits_{\mathbb{E}\left[d(S_{i}, \hat{S}_{i})\right] \leq D_k}I(S_{i}, \hat{S}_{i}),~~~~~\forall i.
\end{equation}

Our goal in this paper is to characterize the \textit{cache capacity-delivery rate tradeoff} for a caching system with $N$ files and $K$ users with arbitrary distortion requirements for both the centralized and decentralized scenarios. 

\begin{Remark}
We note here that, when all the users have the same distortion target, our problem reduces to the classical setting of \cite{maddah2014fundamental} and \cite{maddah2013decentralized}, in which users request files in their entirety. We can consider that files are stored in the server in their compressed form. We also highlight that, in addition to allowing requests at different QoS levels, we extend \cite{maddah2014fundamental} and \cite{maddah2013decentralized} in another direction by allowing different cache capacities at the users.  
\end{Remark}

\begin{Remark}
To illustrate the richness of this model, consider the special case of a library with a single file, and two users, where only one user is equipped with a cache memory. This is equivalent to the well-known successive refinement problem, where the delivery phase corresponds to the base layer, while the placement phase, i.e., the cache content corresponds to the refinement layer.     
\end{Remark}

To further simplify the problem, we will assume that the server employs scalable coding, also known as successive refinement \cite{equitz1991successive}; that is, each source sequence is encoded into multiple layers, each layer targeting a reduced distortion requirement. The base layer contains the least amount of details which enables a lowest-quality reconstruction of each video, while the following layers, referred to as the \textit{enhancement layers}, can refine the quality of reconstruction successively. We remark that an enhancement layer is useful to a user only if this user has all the preceding enhancement layers and the base layer.  

Equivalently, a scalable code maps the required QoS level of each user to a different code layer, where the first layer of each  source sequence, consisting of $r_1$ bits per source sample (bpss) and referred to as \textit{$r_1$-description}, provides an average distortion of $D_1$ when decoded successfully, while the $k$th layer, $k=2, ..., K$, consists of $r_k-r_{k-1}$ bpss, and having received the first $k$ layers, referred to as \textit{$r_k$-description}, a user achieves an average distortion of $D_k$.      
%Following definitions will be instrumental in presenting our main results. Let $D(r)$ denote the \textit{distortion-rate function} of a Gaussian source $S \sim \mathcal{N}(0, \sigma^2)$ encoded at rate $r$ bits per source sample (bpss).  See \cite{cover2012elements} for a rigorous definition of the distortion-rate function. We have $D(r) \triangleq \sigma^2 2^{-2r}$. Let $r_k$ be the minimum compression rate that achieves $D_k$. We have
%\begin{equation}\label{ratedistortion}
%r_k \triangleq R(D_k) = \frac{1}{2}\log_2\frac{\sigma^2}{D_k}, k=1, ..., K.
%\end{equation}
%This means that, to achieve the target distortion of $D_k$, the user has to receive a minimum of $nr_k$ bits corresponding to its desired file. Note that we have $r_1 \leq r_2 \leq \cdots \leq r_K$.

\begin{Remark}
Here we assume for simplicity that all the files in the database have the same distribution, which, in turn, leads to all the compressed files for a fixed distortion target to have the same rate, i.e., the same size. While this greatly simplifies the exposition, extension of the results to the more general scenario with files with different distributions is possible by zero-padding shorter codewords before multicasting, similarly to \cite{ZhangCaching}. 
\end{Remark}

%We will heavily exploit the \textit{successive refinability} of a Gaussian source under squared-error distortion measure~\cite{equitz1991successive}. Successive refinement refers to compressing a sequence of source samples in multiple stages, such that the quality of reconstruction improves, i.e., distortion reduces, at every stage. A given source is said to be successively refinable under a given distortion measure if the single resolution distortion-rate function can be achieved at every stage. Successive refinement has been extensively studied in the source coding literature~\cite{hassanzadeh2015distortion}.

\begin{Remark}
 We remark here that with scalable coding we may not be able to meet the optimal rate-distortion function at each layer, that is, in general, $r^*_k\neq r_k$, since scalable coding may introduce coding overheads to allow various trade-offs among the rates of different layers. However, for some sources and under certain distortion measures scalable coding does not introduce rate loss, called successive refinability. For example, any finite alphabet source with Hamming distortion and Gaussian sources with square-error distortion are examples of successive refinable source and distortion measure pairs~\cite{equitz1991successive, hassanzadeh2015distortion}.
\end{Remark}
\section{Centralized Coded Caching With Distortion Requirements}\label{section:3}

In this section, we investigate the lossy caching problem in the centralized setting. We start by presenting a theoretical lower bound on the delivery rate of the caching system described in Section~\ref{sec1}. We will then consider achievable lossy caching schemes, first for the two-user two-file ($N=K=2$) scenario, then for the general two-user scenario, and finally for an arbitrary number of users and files.

\subsection{Theoretical Lower Bound}

We provide a lower bound on the delivery rate for the general setting ($K$ users and $N$ files) based on the cut-set arguments in Theorem 1.

\begin{Theorem}(Cut-set Bound)
For the lossy caching problem described in Section~\ref{sec1}, the optimal achievable delivery rate is lower bounded by
\begin{align}
&R^\star(M_1, ..., M_K)\nonumber\\
&\geq \operatorname*{max}\limits_{s\in \{1,...,\min\{N, K\}\}} \operatorname*{max}\limits_{\begin{subarray}{c}
    \mathcal{U} \subset \{1,...,K\}\\
    |\mathcal{U}|=s
  \end{subarray}}\left(\sum\limits_{k\in \mathcal{U}} r^*_{k}-\frac{\sum\limits_{k\in\mathcal{U}}M_k}{\lfloor N/s\rfloor}\right). 
\end{align}
\begin{proof}
The proof can be found in Appendix A.
\end{proof}
\end{Theorem}

It is known that the cut-set bound is not tight in general for the centralized coded caching problem~\cite{maddah2014fundamental}. We present another bound for the two-user case ($K=2$), which, together with the cut-set bound, provides a tight lower bound on the delivery rate in certain scenarios. 

\begin{Theorem}
For the lossy caching problem described in Section~\ref{sec1} with $K=2$, we have
\begin{align}
R^\star(M_1, M_2) \geq \frac{r^*_1}{2}+r^*_2-\frac{(M_1+M_2)}{2\lfloor N/2 \rfloor}. 
\end{align}
\begin{proof}
The proof can be found in Appendix B.
\end{proof}
\end{Theorem}

\subsection{Optimal Lossy Caching: Two Users and Two Files $(N=K=2)$}\label{section:2a}

In this section, we characterize the optimal cache capacity-delivery rate tradeoff for the lossy caching problem with two users ($K=2$) and two files ($N=2$), assuming that the underlying source distribution is \textit{successively refinable}. We first present the lower bound on the delivery rate for given $M_1$ and $M_2$ in this particular scenario, followed by the coded caching scheme achieving this lower bound.  

\begin{Corollary}\label{lemma_NK2}
For the lossy caching problem with $N=K=2$, a lower bound on the cache capacity-delivery rate tradeoff is given by
\begin{align}
R^\star (M_1, M_2) \geq  R_c(&M_1, M_2)\triangleq \max\{r^*_1-M_1/2,\nonumber\\ &r^*_2-M_2/2, r^*_1+r^*_2-(M_1+M_2),\nonumber\\ &
r^*_1/2+r^*_2-(M_1+M_2)/2, 0\}. \label{eq444}
\end{align}
\end{Corollary}
The first three terms in RHS of (\ref{eq444}) are derived from the cut-set bound in Theorem 1, and the forth term follows from Theorem 2.
\begin{figure}[t]
\centering
\includegraphics[width=1\linewidth]{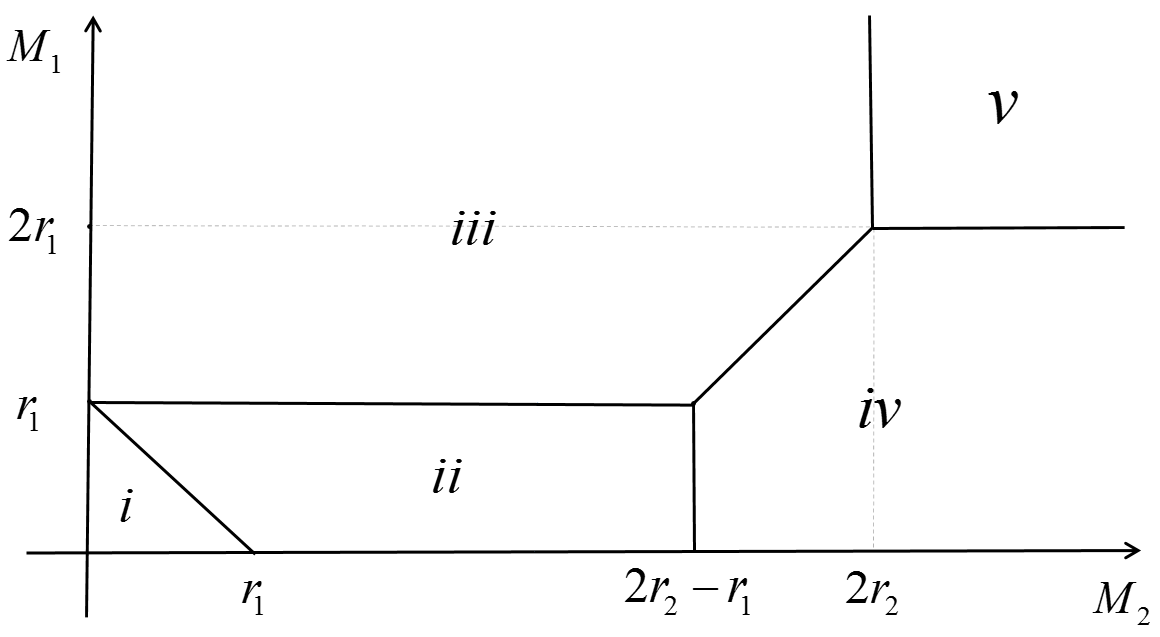}
  \caption{Illustration of the five distinct cases of the cache capacities, $M_1$ and $M_2$, depending on the distortion requirements of the users, $r_1$ and $r_2$ ($N=K=2$).}
\end{figure}

Now we consider the achievability, assuming that a scalable code is employed to compress each file into two layers, where the first layer of rate $r_1$ allows achieving a distortion of $D_1$, while the second layer of rate $r_2-r_1$, together with the first layer results in distortion $D_2$. We denote by $A(B)$ the source codeword of length $nr_2$ bits that can achieve a distortion of $D_2$ for file $S^n_1(S^n_2)$. We refer to the first $nr_1$ bits as the \textit{base layer}, and the remaining $n(r_2-r_1)$ bits as the \textit{refinement layer}. We consider the following five cases depending on the cache capacities of the users, illustrated in Fig.~1: \textit{Case i}, $M_1 + M_2\leq r_1$;\textit{Case ii}, $M_1 + M_2>r_1$, $M_1 \leq r_1$, $M_2 \leq 2r_2-r_1 $; \textit{Case iii}, $M_1>r_1$, $M_2 \leq 2r_2$, $M_2-M_1 \leq 2r_2-2r_1$; \textit{Case iv}, $M_1\leq 2r_1$, $M_2 > 2r_2-r_1$, $M_2-M_1 > 2r_2-2r_1$; \textit{Case v}, $M_1>2r_1$, $M_2>2r_2$. 

%In this case, $R_c(M_1, M_2)=r_1+r_2-(M_1+M_2)$.
%We have $R_c(M_1, M_2)= \frac{r_1}{2}+r_2-\frac{M_1+M_2}{2}$.
%We have $R_c(M_1, M_2)= r_2-\frac{M_2}{2}$.
%We have $R_c(M_1, M_2)=r_1-\frac{M_1}{2}$.
%We have $R_c(M_1, M_2)=0$.

Before presenting the caching and delivery schemes for each of the above cases in detail, we briefly review two different techniques that have been considered in the literature for loseless coded caching systems. Assume that there are two files $W_1$ and $W_2$ in the library, and each of the two users requests one of these files fully. In the placement phase we divide each file into two equal-size non-intersecting sub-files, i.e., we have $W_1=(W_{11}, W_{12})$ and $W_2=(W_{21}, W_{22})$. When the cache sizes of the users are small, the users can cache XORed subfiles. For example, user 1 caches $W_{11}\oplus W_{21}$, while user 2 caches $W_{12}\oplus W_{22}$. Now, consider that user $i$ requests file $W_i$, $i=1, 2$. To satisfy the demands, the server transmits $W_{21}$ and $W_{12}$, which are simultaneously useful to both users. Note that, for cache capacities $M_1=M_2=1/2$, we achieved a delivery rate of $R=1$. This method, referred to as \textit{coded placement}, is first proposed in \cite{maddah2014fundamental} for $N=K=2$, and then generalized by \cite{chen2014fundamental}. Alternatively, if the users have larger caches, e.g., $M_1=M_2=1$, they can cache disjoint subfiles. For example, user 1 caches $W_{11}, W_{21}$, while user 2 caches $W_{12}, W_{22}$. The same demand combination as above can be satisfied by simply transmitting $W_{21} \oplus W_{12}$ to the users which again benefits both users. This method, also proposed by Maddah-Ali and Niesen in~\cite{maddah2014fundamental}, is referred to as \textit{coded delivery}. The optimal caching scheme for the two-user two-file scenario for the lossless caching problem with the same cache sizes, i.e, $M_1=M_2$, has been derived by memory sharing between these two methods and the two extreme cases corresponding to no user cashes ($M_1=M_2=0$) and users caching both files ($M_1=M_2=2$). 

For the lossy caching problem studied in this paper, where the two users have different distortion requirements and different cache sizes, to fully exploit these two methods, we divide the first layers of codewords $A$ and $B$ into six disjoint parts denoted by $A_1$, $\ldots$, $A_6$ and $B_1$, $\ldots$, $B_6$, respectively, where the XORed combination of $A_1$ and $B_1$, i.e., $A_1 \oplus B_1$, is cached by user 1,  the XORed combination of $A_2$ and $B_2$, i.e., $A_2 \oplus B_2$, is cached by user 2, $A_3$ and $B_3$ are cached exclusively by user 1, $A_4$ and $B_4$ are cached exclusively by user 2, $A_5$ and $B_5$ are cached by both user 1 and user 2, and $A_6$ and $B_6$ are cached by none of the users. Since only user 2 requires the refinement layers, we divided the refinement layers of $A$ and $B$ into two disjoint parts denoted by $A_7$, $A_8$ and $B_7$, $B_8$, respectively, where $A_7$ and $B_7$ are cached by user 2, and $A_8$ and $B_8$ are not cached by any user. Overall, user 1 caches $Z_1=\left(A_1\oplus B_1, A_3, B_3, A_5, B_5\right)$, while user 2 caches $Z_2=\left(A_2\oplus B_2, A_4, B_4, A_5, B_5, A_7, B_7\right)$. Let $|A_i|=|B_i|$ for $i=1, ..., 8$, and $|A_3|=|A_4|$ where $|X|$ denotes the length of the binary sequence $X$ (normalized by $n$).
%
%Note that, thanks to the successive refinability of the Gaussian sources, a receiver having received the first $k$ portions, for $k=1, 2, .., 8$, will achieve a distortion level of $D(\sum\limits_{i=1}^k |A_i|)$ for signal $S_1$; and similarly for $S_2$.
\begin{table*}
\label{table1}
\centering
\caption{Illustration of Cache Placement For $N=K=2$}
\begin{tabular}{|l|c|c|c|c|c|c|c|c|}
\hline
       & \multicolumn{6}{c|}{Base Layer} & \multicolumn{2}{c|}{Refinement Layer} \\ \hline
$S_1$  & $A_1$    & $A_2$    & $A_3$   & $A_4$   & $A_5$   & $A_6$   & $A_7$                         & $A_8$                        \\ \hline
$S_2$  & $B_1$    & $B_2$    & $B_3$   & $B_4$   & $B_5$   & $B_6$   & $B_7$                         & $B_8$                                                \\\hlinewd{1.2pt}
Case i  & $M_1$    & $M_2$    & $0$   & $0$   & $0$   & $r_1-M_1-M_2$   & $0$                         & $r_2-r_1$                        \\ \hline
Case ii  & $M_1$    & $r_1-M_1$    & $0$   & $0$   & $0$   & $0$   & $\frac{M_1+M_2-r_1}{2}$                         & $r_2-r_1-\frac{M_1+M_2-r_1}{2}$              \\ \hline
Case iii  & $r_1-l_1-2l_2$    & $0$    & $l_2$   & $l_2$   & $l_1$   & $0$   & $l_3$                         & $r_2-r_1-l_3$              \\ \hline
Case iv  & $0$    & $r_1-M_1$    & $M_1/2$   & $M_1/2$   & $0$   & $0$   & $r_2-r_1$                         & $0$              \\ \hline
Case v  & $0$    & $0$    & $0$   & $0$   & $r_1$   & $0$   & $r_2-r_1$                         & $0$              \\ \hline
\end{tabular}
\end{table*}

Table~I illustrates the placement of contents into users' caches for the five cases by specifying the size of each portion in each case. For example, the forth row implies that in \textit{Case i}, $|A_1|=|B_1|=M_1$, $|A_2|=|B_2|=M_2$,  $|A_6|=|B_6|=r_1-M_1-M_2$, $|A_8|=|B_8|=r_2-r_1$, and the sizes of all other portions are equal to $0$, which is equivalent to dividing $A(B)$ into four portions $A_1(B_1)$, $A_2(B_2)$, $A_6(B_6)$ and $A_8(B_8)$. Thus, in the placement phase, user 1 caches $Z_1=A_1\oplus B_1$, and user 2 caches $Z_2=A_2 \oplus B_2$ so that $|Z_1|=M_1$ and $|Z_2|=M_2$, which meets the cache capacity constraints. For \textit{Case ii}, as presented in the fifth row, $|A_1|=|B_1|=M_1$, $|A_2|=|B_2|=r_1-M_2$,  $|A_7|=|B_7|=\frac{M_1+M_2-r_1}{2}$, $|A_8|=|B_8|=r_2-r_1-\frac{M_1+M_2-r_1}{2}$, and the sizes of all other portions are equal to $0$, which is equivalent to dividing $A(B)$ into four portions $A_1(B_1)$, $A_2(B_2)$, $A_7(B_7)$ and $A_8(B_8)$. Thus, user 1 caches $Z_1=A_1\oplus B_1$, and user 2 caches $Z_2=\{A_2 \oplus B_2, A_7, B_7\}$ so that $|Z_1|=M_1$ and $|Z_2|=M_2$, which meets the cache capacity constraints. The cache placements for the other three cases are presented in a similar manner in Table~I.

Next, we focus on the delivery phase, and identify the minimum delivery rate in each case to satisfy the demands, $\mathbf{d}=(S^n_1, S^n_2)$. Other demand combinations can be satisfied similarly, without requiring higher delivery rates.

\textit{Case i} ($M_1 + M_2\leq r_1$): The server sends $B_1$, $A_2$, $A_6$, $B_6$ and $B_8$. Thus, the delivery rate is $R(M_1, M_2)= r_1+r_2-(M_1+M_2)$.

\textit{Case ii} ($M_1 + M_2>r_1$, $M_1 \leq r_1$, $M_2 \leq 2r_2-r_1 $): The server sends $B_1$, $A_2$ and $B_8$. We have $R(M_1, M_2)= \frac{r_1}{2}+r_2-\frac{M_1+M_2}{2}$.

\textit{Case iii} ($M_1>r_1$, $M_2 \leq 2r_2$, $M_2-M_1 \leq 2r_2-2r_1$): The values of $l_1$, $l_2$ and $l_3$ in Table I are given as:
\begin{subequations}
\begin{equation}
l_1=\max\{0,\min\{M_1-r_1, M_2/2-(r_2-r_1)\}\};\end{equation} \begin{equation}l_2=\max\{0, M_2/2-(r_2-r_1)-l_1\};\end{equation} \begin{equation}l_3=\min\{r_2-r_1, M_2/2\}.\end{equation}
\end{subequations}
The server sends $B_1$, $B_3\oplus A_4$ and $B_8$, which results in
$R(M_1, M_2)= r_2-\frac{M_2}{2}$.

\textit{Case iv}~($M_1\leq 2r_1$, $M_2 > 2r_2-r_1$, $M_2-M_1 > 2r_2-2r_1$): The server sends $B_2$, $B_3\oplus A_4$ and we have
$R(M_1, M_2)=r_1-\frac{M_1}{2}$.

\textit{Case v} ($M_1>2r_1$, $M_2>2r_2$): The cache capacities of both users are sufficient to cache the required descriptions for both files. Thus, any request can be satisfied from the local caches at the desired distortion levels, and we have $R(M_1, M_2)=0$.
\begin{Theorem}\label{theorem_NK2}
For $N=K=2$, and a successively refinable source distribution, we have $R^*(M_1, M_2)=R_c(M_1, M_2)$, i.e., the proposed coded caching scheme meets the lower bound in Corollary \ref{lemma_NK2}; and hence, it is optimal.
\begin{proof}
Theorem~\ref{theorem_NK2} follows by setting $r_1=r^*_1$ and $r_2=r^*_2$ in the above achievable scheme. We note that the achieved delivery rates match those of the lower bound in Corollary~\ref{lemma_NK2} for each of the five cases.
\end{proof}
\end{Theorem}
\begin{Remark}
The optimality of the proposed scheme arises from the maximization of the multicasting opportunities in the discussed five cases. For example, in Case i, by placing disjoint XORed contents from the base layers into the caches of both users, $M_1+M_2$ bits sent in the delivery phase is simultaneously useful to both users regardless of any demand combination. This corresponds to the maximal multicasting opportunities that can be created by any caching scheme as validated by the lower bound on the cache capacity-delivery rate tradeoff given in Corollary 1. 
\end{Remark}
\begin{Remark}
In the special case of identical distortion requirements at the two users, i.e., when $D_1=D_2$, Theorem \ref{theorem_NK2} generalizes the optimal delivery rate result of \cite{maddah2014fundamental} for $N=K=2$ to different cache capacities.  
\end{Remark}

\subsection{Lossy Caching: Two Users and $N$ Files ($K=2, N>2$)}\label{section:2b}

Next, we investigate the more general case with two users and an arbitrary number of files, i.e., $K=2, N>2$. We first present a lower bound on the delivery rate, in Lemma 1, and then present a coded caching scheme, followed by the analysis of the gap between the two.

\begin{Lemma}
\label{lemma_N3K2}
For the lossy caching problem with $K=2$ and $N>2$, a lower bound on the cache capacity-delivery rate tradeoff is given by
\begin{align}\label{eq45}
R^\star &(M_1, M_2) \geq  R_c(M_1, M_2)\triangleq \max\left\{r^*_1-\frac{M_1}{N}, r^*_2-\frac{M_2}{N}, \right.\nonumber \\  
&~\left.\frac{r^*_1}{2}+r^*_2-\frac{M_1+M_2}{2\lfloor N/2 \rfloor}, r^*_1+r^*_2-\frac{M_1+M_2}{2\lfloor N/3 \rfloor}, 0\right\}. 
\end{align}
\end{Lemma}
The first two terms in RHS of (\ref{eq45}) follow from the cut-set bound in Theorem 1, while the third term from Theorem 2. The proof of the forth term can be found in Appendix C.

Next, we present a coded caching scheme for this scenario. Similarly to Section~\ref{section:2a}, we employ scalable coding to compress the files. We denote by $W_j$ the source codeword of length $nr_2$ bits that leads to a distortion of $D_2$ for file $S^n_j$, $j=1, ..., N$. First $nr_1$ bits of $W_j$ corresponds to the base layer that would provide a distortion level of $D_1$ if received. Note that, since $N>K$, \textit{coded placement}, which places XORed contents into users' caches as described in Section III-B, no longer creates multicasting opportunities in this scenario. Hence, we exploit only \textit{coded delivery}, where users cache distinct uncoded subfiles, and XORed subfiles are sent in the delivery phase as described in Section III-B. Accordingly, we divide the base layer of $W_j$, i.e., the first $nr_1$ bits, into four disjoint parts, denoted by $W_{j1}$, $W_{j2}$, $W_{j3}$, and $W_{j4}$, and let user $1$ cache $W_{j1}$ and $W_{j2}$, while user $2$ caches $W_{j2}$ and $W_{j3}$, $j=1, ..., N$. Hence, in the delivery phase, for any demand pair, the server sends $W_{d_13}\bar{\oplus}W_{d_21}$, $W_{d_14}$, $W_{d_24}$, to enable both users to obtain the base layers of their required files. Here $\bar{\oplus}$ represents the bitwise XOR operation, where the arguments are first zero-padded to the length of the longest one. The refinement layer, i.e., the remaining $n(r_2-r_1)$ bits of $W_j$, is divided into two disjoint parts, denoted by $W_{j5}$, $W_{j6}$, $j=1, ..., N$. User 2 caches $W_{j5}$, while user 1 does not cache any part of the refinement layers. In the delivery phase, the server sends $W_{d_26}$ to enable user 2 to obtain the second layer of its requested file. Let the size of each part be the same for all the files, e.g., $|W_{ik}|=|W_{jk}|$, $\forall i, j \in \{1, ..., N\}$, for $k=1, ..., 6$. To conclude, we have $Z_1=\bigcup\limits_{j=1}^N \{W_{j1}, W_{j2}\}$ and $Z_2=\bigcup\limits_{j=1}^N \{W_{j2}, W_{j3}, W_{j5}\}$. By sending $W_{d_13}\bar{\oplus}W_{d_21}$, $W_{d_14}$, $W_{d_24}$, and  $W_{d_26}$, for any possible demand pair $(d_1, d_2)$, in the delivery phase, both users can achieve their target distortion values. 

In the following, we specify the size of each portion depending on the cache capacities $M_1$ and $M_2$, and derive the corresponding delivery rate. We note that the multicasting opportunities only come from $W_{j1}$ and $W_{j3}$ (by sending $W_{d_13}\bar{\oplus}W_{d_21}$ in the delivery phase as described above). Hence, our goal is to maximize $W_{d_13}\bar{\oplus}W_{d_21}$ in order to minimize the delivery rate. We consider three cases:

\textit{Case i} ($M_1 + M_2\leq Nr_1$): We let: $|W_{j1}|=M_1/N$; $|W_{j2}|=0$; $|W_{j3}|=M_2/N$; $|W_{j4}|=r_1-M_1/N-M_2/N$; $|W_{j5}|=0$; $|W_{j6}|=r_2-r_1$.  The delivery rate is $R(M_1, M_2)= r_1+r_2-\frac{2(M_1+M_2)}{N}+\frac{\max\{M_1, M_2\}}{N}$.

\textit{Case ii} ($M_1 + M_2 > Nr_1$, $M_1 \leq Nr_1$): We let: $|W_{j1}|=\max\{\min\{r_2-M_2/N, M_1/N\}, 0\}$; $|W_{j2}|=\max\{M_1/N+M_2/N-r_2,0\}$; $|W_{j3}|=r_1-M_1/N$; $|W_{j4}|=0$; $|W_{j5}|=\min\{M_1/N+M_2/N-r_1, r_2-r_1\}$; $|W_{j6}|=\max\{r_2-(M_1/N+M_2/N), 0\}$. We have $R(M_1, M_2)= \max\{r_2-(M_1+M_2)/N, 0\}+\max\{\min\{r_2-M_2/N, M_1/N\}, r_1-M_1/N\}$.

\textit{Case iii} ($M_1 > Nr_1$): Let $|W_{j1}|=r_1-\min\{M_2/N, r_1\}$; $|W_{j2}|=\min\{M_2/N, r_1\}$; $|W_{j3}|=0$; $|W_{j4}|=0$; $|W_{j5}|=\max\{0, \min\{r_2, M_2/N\}-r_1\}$; $|W_{j6}|=\min\{r_2-r_1, \max\{0, r_2-M_2/N\}\}$. It yields 
$R(M_1, M_2)= \max\{0, r_2-M_2/N\}$.

We can summarize the achievable delivery rate as follows:

\begin{align}\label{eq44}
&R(M_1, M_2)\nonumber\\
&=\begin{cases}
r_1+r_2-2M_1/N-M_2/N, &\mbox{if}~~(M_1, M_2)\in \mathcal{M}_1,\\
r_1+r_2-M_1/N-2M_2/N, &\mbox{if}~~(M_1, M_2)\in \mathcal{M}_2,\\
r_1-M_1/N, &\mbox{if}~~(M_1, M_2)\in \mathcal{M}_3,\\
r_2-M_2/N, &\mbox{if}~~(M_1, M_2)\in \mathcal{M}_4,\\
0, &\mbox{if}~~(M_1, M_2)\in \mathcal{M}_5,
\end{cases}
\end{align}
where $\mathcal{M}_1$, $\mathcal{M}_2$, $\mathcal{M}_3$, $\mathcal{M}_4$ and $\mathcal{M}_5$, illustrated in Fig. 2, are specified as follows: 
\begin{figure}[t!]
\centering
    \includegraphics[width=1.0\linewidth]{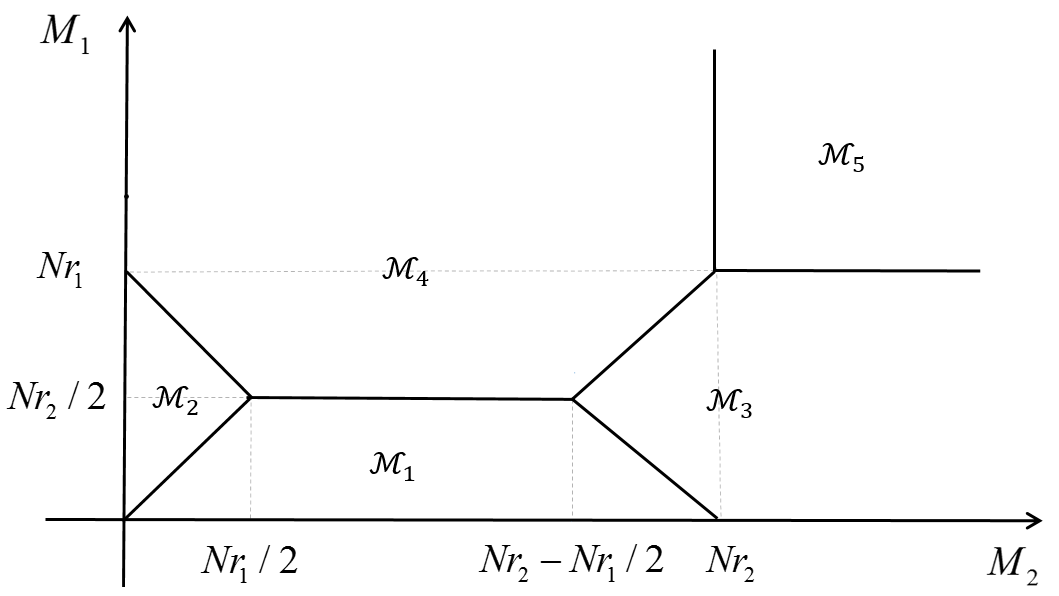}
    \caption{Illustration of the five distinct cases of the cache capacities, $M_1$ and $M_2$, depending on the distortion requirements of the users, $r_1$ and $r_2$ ($K=2, N>2$).}
\end{figure}
\begin{subequations}
\begin{align}
\mathcal{M}_1=&\{(M_1, M_2) | 0 \leq M_1 \leq Nr_1/2, M_1 + M_2\leq Nr_2,\nonumber\\
&\qquad\qquad\qquad\qquad\qquad\qquad\qquad M_1 \leq M_2\};\\
\mathcal{M}_2=&\{(M_1, M_2) | M_1 + M_2\leq Nr_1, M_1> M_2, M_2\geq 0\};\\
\mathcal{M}_3=&\{(M_1, M_2)|0 \leq M_1 \leq Nr_1, M_1 + M_2> Nr_2, \nonumber\\
&\qquad\qquad\qquad\qquad M_2-M_1> N(r_2-r_1)\};\\
\mathcal{M}_4=&\{(M_1, M_2)|M_1>Nr_1/2, 0 \leq M_2 \leq Nr_2,\nonumber\\
&~~M_1 + M_2>Nr_1, M_2-M_1\leq N(r_2-r_1)\};\\
\mathcal{M}_5=&\{(M_1, M_2) | M_1 >Nr_1, M_2>Nr_2\}.
\end{align}
\end{subequations}

If the underlying source distribution is successively refinable, i.e., $r_1=r^*_1$ and $r_2=r^*_2$, comparing (\ref{eq45}) with (\ref{eq44}), we can conclude that for $(M_1, M_2) \in \mathcal{M}_3 \bigcup \mathcal{M}_4 \bigcup \mathcal{M}_5$, the achieved delivery rate meets the lower bound in Lemma 1. We analyse the optimality of the achieved delivery rate for $(M_1, M_2) \in \mathcal{M}_1 \bigcup \mathcal{M}_2$ in the following, assuming successive refinability, i.e., $r_1=r^*_1$ and $r_2=r^*_2$. Consider three cases: $N=3c+i$, for $i=0, 1, 2$ and $c \in \mathbb{Z}^+$.

\textit{Case 1} ($N=3c$): The last term of (\ref{eq45}) can be rewritten as $r^*_1+r^*_2-\frac{3(M_1+M_2)}{2N}$.Thus, $\Delta R \triangleq R(M_1, M_2)-R_c(M_1, M_2) = \left|\frac{M_1-M_2}{2N}\right|$for any $(M_1, M_2)\in \mathcal{M}_1 \bigcup \mathcal{M}_2$. Note that when $M_1=M_2$, $\Delta R=0$, that is, the achieved delivery rate meets the lower bound.

\textit{Case 2} ($N=3c+1$): We can rewrite the last term of (\ref{eq45}) as $r^*_1+r^*_2-\frac{3(M_1+M_2)}{2(N-1)}$. We have $\Delta R = \left|\frac{M_1-M_2}{2(N-1)}\right|+\frac{\max\{2M_1+M_2, 2M_2+M_1\}}{N(N-1)}$, for any $(M_1, M_2)\in \mathcal{M}_1 \bigcup \mathcal{M}_2$. 

\textit{Case 3} ($N=3c+2$): The last term of (\ref{eq45}) equals to  $r_1+r_2-\frac{3(M_1+M_2)}{2(N-2)}$. We have $\Delta R = \left|\frac{M_1-M_2}{2(N-2)}\right|+\frac{2\max\{2M_1+M_2, M_1+2M_2\}}{N(N-2)}$, for any $(M_1, M_2)\in \mathcal{M}_1 \bigcup \mathcal{M}_2$. 

\begin{Corollary}
For $K=2$, when $N$ is divisible by $3$ and the cache capacities of the users are identical, i.e., $M_1=M_2$, if the underlying source is successively refinable, the proposed coded caching scheme meets the lower bound in Lemma \ref{lemma_N3K2}; and hence, it is optimal, i.e., $R^*(M_1, M_2)=R_c(M_1, M_2)$.
\end{Corollary}

\begin{Remark}
When $D_1=D_2$, Corollary 2 generalizes the optimal delivery rate result of \cite{maddah2014fundamental} for $N=K=2$ to all $N$ values that are multiples of $3$. 
\end{Remark}

\subsection{General Case}\label{sec2c}

In this section, we tackle the general scenario with $N$ files and $K$ users. As before, we denote by $r_k$ the compression rate by the used scalable code that achieves an average distortion of $D_k$, $k=1, ..., K$, where $D_1 \geq D_2 \geq \cdots \geq D_K$. We remind the reader that scalable code provides a layered structure of descriptions for each file, where the first layer consists of $nr_1$ bits, and achieves distortion $D_1$ when decoded successfully. The $k$th layer, $k=2, ..., K$, consists of $n(r_{k}-r_{k-1})$ bits, and having received the first $k$ layers, a user achieves a distortion of $D_k$.

The examples studied in Sections~\ref{section:2a} and~\ref{section:2b} illustrate the complexity of the lossy content caching problem; we had to consider five cases even with two users and two files. The problem becomes intractable quickly with the increasing number of files and users. However, note that, only users $k, k+1, ..., K$, whose distortion requirements are lower than $D_k$, need to decode the $k$th layer for the file they request, for $k=1, ..., K$. Therefore, once all the contents are compressed into $K$ layers based on the distortion requirements of the users employing scalable code, we have, for each layer, a lossless caching problem. However, each user also has to decide how much of its cache capacity to allocate for each layer. Hence, the lossy caching problem under consideration can be divided into two subproblems: the lossless caching problem of each source coding layer, and the cache allocation problem among different layers.

In general, the optimal delivery rate achieved by this layered algorithm can be found by jointly optimizing the cache capacity allocated by each user to each layer, and the caching and delivery scheme to be used for each layer to minimize the corresponding delivery rate. However, we first note that the optimal delivery rate remains an open problem even in the case of equal cache capacities. Therefore, we are bound to use achievability results. Moreover, these achievability results are often characterized as piecewise linear functions, and memory-sharing among multiple schemes may be required, further complicating the identification of the minimal delivery rate. Therefore, we are not able to provide a low complexity algorithm that can optimize all the system parameters jointly. Indeed, we will propose heuristic cache allocation schemes at the users, and provide a suboptimal caching and delivery algorithm for given cache capacities. 

\subsubsection{Coded Lossless Caching of Each Layer}

We first assume that the cache allocation at the users for each layer is already fixed, and focus on the first subproblem of centralized lossless caching with heterogeneous cache sizes. This problem has previously been studied in \cite{wang2015heterogeneouscachesizes} and \cite{Amiri2016heterogeneouscachesizes} in the decentralized setting, while, to the best of our knowledge, it has not been considered in the centralized setting. Consider, for example, the $k$th refinement layers of all the files. There are only $L_k\triangleq K-k+1$ users (users $k, k+1, ..., K$) who may request these layers. Let user $j$, $j\in \{k, ..., K\}$, allocate $M_{j, k}$ (normalized by $n$) of its cache capacity for this layer. Without loss of generality, we order users $k, ..., K$ according to the cache capacities they allocate, and re-index them, such that $M_{k, k}\leq M_{k+1, k}\leq \cdots \leq M_{K, k}$.

We would like to have symmetry among allocated cache capacities to enable multicasting to a group of users. Based on this intuition, we further divide layer $k$ into $L_k$ sub-layers, and let each user in $\{k, ..., K\}$ allocate $M_{k}^1=M_{k, k}$ of its cache for the first sub-layer, and each user in  $\{k+i-1, ..., K\}$ allocate $M_{k}^i=M_{k+i-1, k}-M_{k+i-2, k}$ of its cache for the $i$th sublayer, for $i=2,\ldots, L_k$. Overall, we have $L_k$ sub-layers, and users $k+i-1, k+i, ..., K$ allocate $M_k^i$ of their caches for sub-layer $i$, whereas no cache is allocated by users $k, k+1, ..., k+i-2$.

We denote by $r_k^i$ the size of the $i$th sub-layer of the $k$th refinement layer, and by $R(L_k, i, M_{k}^i, r_k^i,$ $ N)$ the minimum required delivery rate for this sub-layer. The rates, $r_k^i$, $i=1, ..., L_k$, should be optimized jointly in order to minimize the total delivery rate for the $k$th layer. The optimization problem can be formulated as follows:

\begin{subequations}\label{eq:optim}
\begin{equation}\min \limits_{r_k^1, ..., r_k^{L_k}} \sum_{i = 1}^{L_k} R(L_k, i, M_{k}^i, r_k^i, N)\end{equation}
\begin{equation}\mathrm{s. t.}  \sum_{i = 1}^{L_k} r_k^i=r_k-r_{k-1}.\end{equation}
\end{subequations}

We consider achievable $R(L_k, i, M_{k}^i, r_k^i, N)$ values based on \textit{coded delivery} and \textit{coded placement}, outlined in Section III-B, as well as the group-based centralized caching (GBC) scheme proposed in \cite{mohammadqian2016large}. In \textit{GBC} the user is grouped according to their demands in the delivery phase, i.e., users with the same demand are placed in the same group, and then contents are first exchanged within the same group and then across different groups.For the lossless caching problem, \textit{GBC} is shown to achieve a lower delivery rate compared to memory-sharing between the \textit{coded delivery} and \textit{coded placement} schemes, when each user has a cache capacity of $N/K$ (normalized by the file size). We consider two cases:

Case 1) $L_k < N$. In this case, in the worst case when users $\{k, ..., K\}$ request distinct files, \textit{GBC} and \textit{coded placement} provide no caching gain; thus, we employ \textit{coded delivery}. Focus on the $i$th sub-layer: users $k+i-1, ..., K$ each allocate $M_{k}^i$ of cache capacity, while users $k$ to $k+i-2$ allocate no cache for this sublayer. If $r_k^i \in \mathcal{P}_{MAN}\triangleq \{0, M_{k}^i/N, M_{k}^iL_k^i/((L_k^i-1)N), M_{k}^iL_k^i/((L_k^i-2)N), ..., M_{k}^iL_k^i/N\}$, where $L_k^i=L_k+1-i$,  we have
\begin{align} \label{eq3}
R_{MAN}(&L_k, i, M_{k}^i, r_k^i, N)=(i-1)\cdot r_k^i\nonumber\\
&+r_k^iL_k^i\cdot(1- M_{k}^i/r_k^iN)\cdot\frac{1}{1+M_{k}^iL_k^i/r_k^iN}.
\end{align}
The first term on the right hand side is due to unicasting to users $k$ to $k+i-2$, while the second term is the \textit{coded delivery} rate to users $k+i-1$ to $K$ given in \cite{maddah2014fundamental}. Based on the memory sharing argument, any point on the line connecting two points, $(r_1', R(L_k, i, M_{k}^i, r_1', N))$ and $(r_2', R(L_k, i, M_{k}^i, r_2', N))$, is also achievable, i.e., if $r_k^i \in [0, M_{k}^iL_k^i/N]$ and $r_k^i \notin \mathcal{P}_{MAN}$, then we have
\begin{align} \label{eq4}
R(L_k, &i, M_{k}^i, r_k^i , N)\nonumber\\
=&\min\limits_{\begin{subarray}{c}
   r_1 \in \mathcal{P}_{MAN},~r_1 < r_k^i\\
    r_2 \in \mathcal{P}_{MAN},~r_1 > r_k^i\end{subarray}}\left\{\frac{r_k^i-r_1'}{r_2'-r_1'}R(K_k, i, M_{k}^i, r_1', N)\right.\nonumber\\
&\qquad\qquad\quad\left.+\frac{r_2'-r_k^i}{r_2'-r_1'}R(K_k, i, M_{k}^i, r_2', N)\right\},
\end{align}
and if $r_k^i > M_{k}^iL_k^i/N$, we have
\begin{align} \label{eq5}
R(L_k, i, M_{k}^i, r_k^i, N)=&(i-1)\cdot r_k^i+\frac{M_{k}^iL_k(L_k-1)}{2N}\nonumber\\
&+r_k^iM_{k}^iL_kL_k^i/N.
\end{align}

Case 2) $L_k \geq  N$. In this case, \textit{coded placement} achieves a lower delivery rate than \textit{coded delivery} when $r_k^i\geq M_{k}^iL_k^i$ \cite{chen2014fundamental}. If $L_k^i>N\geq 3$, \textit{GBC} outperforms \textit{coded delivery} at point $r_k^i= M_k^iL_k^i/N$ \cite{mohammadqian2016large}. Note that for the $i$th sub-layer, there are $i-1$ users with no cache allocation. If $i-1 \geq N$, there will be no gain with any of the schemes. When $i-1 < N$ and $r_k^i \geq M_{k}^iL_k^i$, the delivery rate of \textit{coded placement} is
\begin{equation} \label{eq6}
R_{CFL}(L_k, i, M_{k}^i, r_k^i, N)=Nr_k^i-(N-i+1)M_{k}^i.
\end{equation}
If $i-1 < N$, $L_k^i>N\geq 3$ and $r_k^i = M_{k}^iL_k^i/N$, the delivery rate provided by \textit{GBC} is given by 
\begin{align} \label{eq7}
R_{GBC}(L_k, i, M_{k}^i, r_k^i, N)=&(i-1)\cdot r_k^i
+Nr_k^i\nonumber\\
&-\frac{N(N+1)r_k^i}{2L_k^i}.
\end{align}

When $0\leq r_k^i\leq M_{k}^iL_k^i$, the delivery rate is given by the lower convex envelope of points $(M_{k}^iL_k^i, R_{CFL}(L_k, i, M_{k}^i, M_{k}^iL_k^i, N))$ given by (\ref{eq6}), $(M_{k}^iL_k^i/N, R_{GBC}(L_k, i, M_{k}^i, M_{k}^iL_k^i/N, N))$ given by (\ref{eq7}), and $(r_k^i, R_{MAN}(L_k, i, M_{k}^i, r_k^i, N))$; and for $r_k^i \in \mathcal{P}_{MAN}\setminus\{M_k^iL_k^i/N\}$, given by (\ref{eq3}).

\theoremstyle{definition}
\newtheorem{exmp}{Example}

\begin{exmp}
Consider $N=2$ files and $K=3$ users. We focus on the caching and delivery of the first layers by applying the above coded lossless caching scheme. Note that all three users require the base layer of their requested file, such that $L_1=3$. Assume that $r_1=3$ and the allocated cache capacities for this layer by the users are given by $M_{1, 1}=0.5$, $M_{2, 1}=1$, $M_{3, 1}=1.5$, respectively. The first layer of each file is thus divided into $L_1=3$ disjoint sub-layers. All the three users allocate a cache capacity of $M_{1}^1=0.5$ to cache the first sub-layers; user 2 and user 3 allocate a cache capacity of $M_{1}^2=0.5$ to cache the second sub-layers; and user 3 allocates $M_{1}^3=0.5$ of its cache capacity to cache the third sub-layers. The rates of these three sub-layers, $r_1^1, r_1^2, r_1^3$, are optimized in order to minimize the total delivery rate as given by \eqref{eq:optim}. By solving \eqref{eq:optim} numerically, we obtain the optimal partition of the first layer given by $r_1^1=1.5$, $r_1^2=1.0$, and $r_1^3=0.5$. The delivery rates of these three sub-layers are $2$, $1$, $1$, respectively, which results in a total delivery rate of $4$.
\end{exmp}

\subsubsection{Allocation of Cache Capacities}

Next, we propose two algorithms for cache allocation among layers: \textit{proportional cache allocation} (PCA) and \textit{ordered cache allocation} (OCA), which are elaborated in Algorithms~1 and~2, respectively, where $r_k$ is as defined earlier, and $r_0=0$.

\begin{algorithm}\label{alg33}
\caption{Proportional Cache Allocation (PCA)}
\begin{algorithmic}[1]
\State{\textbf{Require:} $\mathbf{r}={r_1, ..., r_K}$}
\For{$k=1, ..., K$}
\For{$i=1, ..., k$}
\State{user $k$ allocates $\frac{r_i-r_{i-1}}{r_k}M_k$ to layer $i$}
\EndFor
\EndFor
\end{algorithmic}
\end{algorithm}

\begin{algorithm}\label{alg44}
\caption{Ordered Cache Allocation (OCA)}
\begin{algorithmic}[1]
\State{\textbf{Require:} $\mathbf{r}={r_1, ..., r_K}$}
\For{$k=1, ..., K$}
\State{user $k$ allocates all of its cache to the first $i$ layers, where $r_{i-1} < \frac{M_k}{N} \leq r_{i}$}
\EndFor
\end{algorithmic}
\end{algorithm}

PCA allocates each user's cache among the layers it may request proportionally to the sizes of the layers, while OCA gives priority to lower layers. The server can choose the one resulting in a lower delivery rate. Numerical comparison of these two cache allocation schemes together with the delivery scheme proposed above will be presented in Section V.

\begin{Remark}
As stated earlier, when the QoS requirements of the users are identical, i.e., $D_1=D_2=\cdots=D_N$, the lossy caching problem considered here is equivalent to the loseless coded caching problem with distinct cache capacities. Therefore, the centralized coded caching scheme proposed in this section is also the first centralized caching scheme that generalizes the centralized caching problem in \cite{maddah2014fundamental} to heterogeneous cache capacities.
\end{Remark}
%Consider the lossy caching problem with $N$ successive refinable sources each of $n$ samples, and $K$ users each having a cache of size $M_kn$ bits, $k=1, ..., K$ and ordered distortion requirements $\mathcal{D}=(D_1, ..., D_K)$ that $D_1 \geq D_2 \geq \cdots \leq D_K$, let $r_k$ be the minimum compression rate that achieves $D_k$,

\section{Decentralized Coded Caching With Distortion Requirements}
In this section, we consider the lossy coded caching problem in the \textit{decentralized setting}; that is, the server is assumed to have only the set of possible distortion values that may be requested by the users, but has no prior knowledge of the number of users and their target distortion requirements. Accordingly, in decentralized caching, the \emph{placement phase} is conducted locally and independently for each user since coordination among users is not possible. We note that the server needs to know the possible distortion values that can be requested by the users in advance in order to compress the files in the database using an appropriate successive refinement source code.

In the \textit{placement phase} of the proposed coded caching scheme, \textit{user k randomly caches $M_kn/N$ bits of the $r_k$-description of each file, for $k=1, ..., K$.} Recall that the $r_k$-description corresponds to the codeword of length $nr_k$ bits by scalable code, which achieves an average distortion of $D_k$ when decoded. As mentioned in the previous section, scalable code provides a layered structure of descriptions for each file; that is, $r_k$-distortion corresponds to the first $nr_k$ bits of the $r_{k+1}$-description. Different layers of  each file are cached by a different subset of users, and cached contents of each layer occupy different sizes of memory due to heterogeneous distortion requirements and heterogeneous cache sizes. To illustrate the achievable delivery rate, we first present an example with two users and two files $(K=N=2)$, and then extend our analysis to the general scenario.

\begin{Remark}
Here we assume that the server has the knowledge of all possible distortion values that may be requested by the users. This is needed for the server to employ scalable coding with the desired number of layers. Note that the server does not know either the number or the identity of the active users in advance; however, in practice, the number of layers, or equivalently, the number of QoS levels that can be requested will be limited either by the compression scheme employed, or due to the limited variety of devices available, and will be much smaller than the number of users in the system.  
\end{Remark}

\subsection{Two Users and Two Files $(K=N=2)$}
Here we consider the same system model as in Section~\ref{section:2a} with two users and two files $(N=K=2)$. User 1 has a cache of size $M_1$, while user 2 has a cache of size $M_2$. 

In the \textit{placement phase}, user 1 randomly caches $M_1n/2$ bits from the $r_1$-description of each file; while user 2 randomly caches $M_2n/2$ bits from the $r_2$-description of each file. We denote by $A^i_{\mathcal{U}}$ $(B^i_{\mathcal{U}})$ the bits of the $i$th layer of file $S_1$ $(S_2)$ that are cached exclusively by the subset $\mathcal{U}$, where $\mathcal{U}\subset\{1, 2\}$ and $i=1, 2$. For example, $A^1_{1,2}$ denotes the part of the base layer of file $S^n_1$ cached by both users, while $B^2_{2}$ is the segment of bits from the second refinement layer of file  $S^n_2$ cached only by user 2. We list the expected size of each segment in Table~\ref{table2} (normalized by $n$), where $t_i$ denotes the probability of any bit from the $r_i$-description of each file cached by user $i$. We have $t_i=\min\{1, \frac{M_i}{2r_i}\}$, $i=1, 2$.
\begin{table*}
\centering
\caption{Illustration of Cache Placement}
\label{table2}
\begin{tabular}{|l|l|l|l|l|l|l|l|l|}
\hline
& \multicolumn{4}{|l|}{First layer} & \multicolumn{2}{l|}{Second layer} \\ \hline
Segment & $A^1_{\emptyset}(B^1_{\emptyset})$ &   $A^1_{1}(B^1_{1})$  &   $A^1_{2}(B^1_{2})$  &   $A^2_{1,2}(B^2_{1,2})$  & $A^2_{\emptyset}(B^2_{\emptyset})$ &   $A^2_{2}(B^2_{2})$   \\ \hline
Size &$r_1(1-t_1)(1-t_2)$&$r_1t_1(1-t_2)$&$r_1t_2(1-t_1)$& $r_1t_1t_2$&$(r_2-r_1)(1-t_2)$&$t_2(r_2-r_1)$     \\ \hline
\end{tabular}
\end{table*}

In the \textit{delivery phase}, for demand pair $(S^n_1, S^n_2)$, the server sends $A^1_{\emptyset}$, $B^1_{\emptyset}$, $B^2_{\emptyset}$ and $A^1_{2} \bar{\oplus} B^1_{1}$ to satisfy both requests. Note that $A^1_{2}$ and $B^1_{1}$ may be of different sizes. We employ the bitwise XOR operation $\bar{\oplus}$. Any other demand combination can be satisfied in a similar manner. Hence, the worst-case delivery rate is given by
\begin{align}
R=&2r_1(1-t_1)(1-t_2)+(r_2-r_1)(1-t_2)\nonumber\\
&+\max\{r_1t_1(1-t_2), r_1t_2(1-t_1)\}.
\end{align}

\begin{figure}\label{decenN2K2}
\centering
\includegraphics[width=1.05\linewidth]{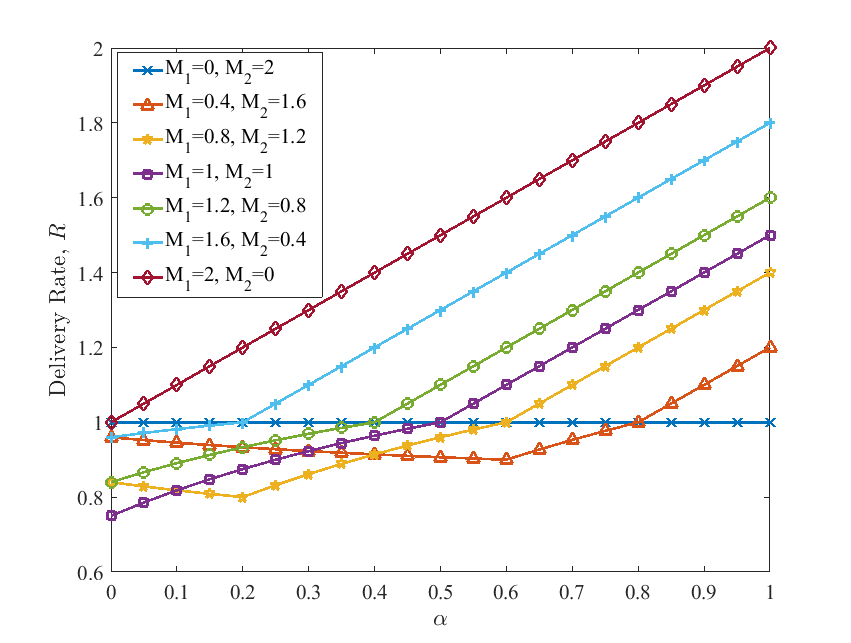}
\caption{Achievable delivery rate for $r_1=1-\alpha$ and $r_2=1+\alpha$, $\alpha \in [0, 1]$, $N=2$, $K=2$.}
\end{figure}

Similarly to centralized coded caching, multicasting is possible only for the delivery of the first layer that is requested by both users. In Fig.~3, we plot the achievable delivery rate for different cache capacity pairs $(M_1, M_2)$, and different distortion requirement pairs $(D_1, D_2)$, such that $r_1=1-\alpha$ and $r_2=1+\alpha$. The value of $\alpha$ varies from $0$ to $1$. The difference between the distortion requirements of the users becomes more significant as $\alpha$ increases, while the sum of the requested rates remains the same. When $\alpha=0$, the users have the same distortion target, and the delivery rate increases with the difference between the cache capacities even though the total cache capacity remains the same. We observe that the pair $(M_1, M_2)$ that achieves the minimum delivery rate at a certain $\alpha$ has the closest ratio between $M_1$ and $M_2$ to the ratio between $r_1$ and $r_2$. For instance, when $\alpha=0.2$, i.e., $r_1=0.8$ and $r_2=1.2$, $(M_1, M_2)=(0.8, 1.2)$ minimizes the delivery rate among the eight cache memory size pairs considered. Similarly, it can be observed that the delivery rate is smaller if the user with the lower distortion requirement has a larger cache capacity. We also observe that $(M_1, M_2)=(2,0)$ always has the highest delivery rate since only one user has cache capacity. Finally, We observe from the $(M_1, M_2)=(1,1)$ curve that, when the two users have the same cache capacity, a larger $\alpha$, i.e., users with more divergent QoS requirements, results in a higher delivery rate. 

\subsection{General Case}
\begin{algorithm}\label{alg3}
\caption{Decentralized Coded Caching Scheme with Lossy Distortion Requests}
\begin{algorithmic}[1]
\Statex
\Procedure{PLACEMENT}{}
\For{$k=1, ..., K$}
\For{$j=1, ..., N$}
\State{User $k$ randomly caches $M_kn/N$ bits of the $r_k$-description of file $j$}
\EndFor
\EndFor
\EndProcedure

\Procedure{DELIVERY1 $(d_1, ..., d_K)$}{}
\For{$k=1, ..., K$}
\For{$\mathcal{S} \subset \{1, ..., K\}$: $|\mathcal{S}|=k$}
\State{Send $X_{\mathcal{S}}=\bar{\oplus}_{s\in \mathcal{S}}W_{(d_s, D_s),\mathcal{S}\setminus\{s\}}$}
\EndFor
\EndFor
\EndProcedure
\Procedure {DELIVERY2 $(d_1, ..., d_K)$}{}
\For{$i = 1, \ldots, N$}
\State {server sends enough random linear combinations of the bits of the compressed version of file $S^n_{i}$ to enable all the users demanding it to decode it at their desired distortion levels.}
\EndFor
\EndProcedure
\end{algorithmic}
\end{algorithm}

We present a decentralized coded caching scheme for the general scenario in Algorithm~3, based on the decentralized caching scheme of \cite{maddah2013decentralized}, where the parameters $r_k$ are as defined in Section~\ref{sec2c}, and $W_{(d_s, D_s),\mathcal{S}\setminus\{s\}}$ denotes the part of the description of the file $d_s$ requested by user $s$ at distortion level $D_s$, that is cached exclusively by the subset of users $\mathcal{S}\setminus\{s\}$. Algorithm~3 contains two delivery procedures, DELIVERY1 and DELIVERY2, and according to the information on the active users received at the beginning of the delivery phase, the server can choose the delivery procedure with a lower delivery rate. The following lemma provides the expected size of each segment $W_{(d_s, D_s),\mathcal{S}\setminus\{s\}}$.

\begin{Lemma}According to the placement phase of Algorithm~3, as the blocklength $n$ goes to infinity, by the law of large numbers, we have 
\begin{align}\label{eq:4-1}
\frac{|W_{(d_s, D_s),\mathcal{S}\setminus\{s\}}|}{n}\rightarrow \sum\limits_{i=1}^{\inf \mathcal{S}} (r_{i}-r_{i-1})p^i_{(d_s, D_s),\mathcal{S}\setminus\{s\}},
\end{align}
with probability $1$, where $|W_{(d_s, D_s),\mathcal{S}\setminus\{s\}}|$ denotes the size of segment $W_{(d_s, D_s),\mathcal{S}\setminus\{s\}}$. We also have $r_0=0$, and
\begin{align}
p^i_{(d_s, D_s),\mathcal{S}\setminus\{s\}}=&\left(1-\frac{M_s}{Nr_s}\right)\prod\limits_{u\in\mathcal{S}\setminus\{s\}}\frac{M_u}{Nr_u}\cdot\nonumber\\
&\qquad\qquad\prod\limits_{u\in\{i, ..., K\}\setminus\mathcal{S}}\left(1-\frac{M_u}{Nr_u}\right).
\end{align}
\begin{proof}
Based on Algorithm~3, user $k$ caches $M_kn/N$ bits from the $r_k$-description of each file, which implies that each bit of the $r_k$-description is cached by user $k$ with probability $M_k/Nr_k$, and no bit from the higher layers is cached by user $k$. We define $m(\mathcal{S})$ as the user with the smallest index in set $\mathcal{S}$, i.e., $m(\mathcal{S})=\min\{s: s\in \mathcal{S}\}$. Then, each bit in $r_{m(\mathcal{S})}$-description is cached by user $u$, $u\in \mathcal{S}$, with probability $M_u/Nr_u$. Since $D_1 \geq D_2\geq \cdots \geq D_K$, we have $D_{m( \mathcal{S})} \geq D_{u}$, i.e., $r_{m( \mathcal{S})} \leq r_{u}$, for any $u \in \mathcal{S}$. The file segment $W_{(d_s, D_s),\mathcal{S}\setminus\{s\}}$ contains bits only from the $r_{m(\mathcal{S})}$-description of file $S^n_{d_s}$.

Note that user $k$ caches bits only from the first $k$ layers, i.e., the $k$th layer is exclusively cached by users $k, k+1, ..., K$. For the $i$th layer, $i=1, ..., m(\mathcal{S})$, which is cached by users $i, ..., K$, every bit in the $i$th layer of file $d_s$ is exclusively cached by users in subset $\mathcal{S}\setminus\{s\}$ with probability $p^i_{(d_s, D_s),\mathcal{S}\setminus\{s\}}$. Since the total rate of the $i$th layer is $r_i-r_{i-1}$, the expected number of bits from the $i$th layer in $W_{(d_s, D_s),\mathcal{S}\setminus\{s\}}$ is $ n(r_{i}-r_{i-1})p^i_{(d_s, D_s),\mathcal{S}\setminus\{s\}}$. As $W_{(d_s, D_s),\mathcal{S}\setminus\{s\}}$ has bits from the first $m(\mathcal{S})$ layers, we sum up the expected rate of all these layers, which yields (\ref{eq:4-1}).
\end{proof}
\end{Lemma}

Since $\frac{M_k}{Nr_k}$ are not identical for $k=1, .., K$, in the delivery phase, for the multicast subset $\mathcal{S}$, the sizes of the corresponding segments, $W_{(d_s, D_s),\mathcal{S}\setminus\{s\}}$, $s\in\mathcal{S}$, will be different. Therefore, we apply the $\bar{\oplus}$ operation. Hence, the size of the multicasted segment for subset $\mathcal{S}$ is given by
\begin{subequations}\label{eq:4-2}
\begin{align}
|X_{\mathcal{S}}|&=\max\limits_{s\in\mathcal{S}}|W_{(d_s, D_s),\mathcal{S}\setminus\{s\}}|\\
&=|W_{(d_{r'(\mathcal{S})}, D_{ r'(\mathcal{S})}),\mathcal{S}\setminus\{ r'(\mathcal{S})\}}|\\
&=n\sum\limits_{i=1}^{m( \mathcal{S})} (r_{i}-r_{i-1})p^i_{(d_{r'(\mathcal{S})}, D_{r'(\mathcal{S})}),\mathcal{S}\setminus\{r'(\mathcal{S})\}}.
\end{align}
\end{subequations}
where $ r'(\mathcal{S})=\argmin\limits_{s\in\mathcal{S}}\frac{M_s}{Nr_s}$, which is the index of the user in subset $\mathcal{S}$ with the smallest $\frac{M_s}{Nr_s}$.
\begin{algorithm}\label{alg4}
\caption{Layered Content Delivery 1 (LCD1)}
\begin{algorithmic}[1]
\Statex
\Procedure {DELIVER $(d_1, ..., d_K)$}{}
\For {$i=1, ..., K$}
\Procedure {delivery $i$th layer of files $(d_i, ..., d_K)$ to users $\{i, ..., K\}$}{}
\For {$k=1, ..., K+1-i$}
\For {$\mathcal{S} \subset \{i, ..., K\}$: $|\mathcal{S}|=k$}
\State{Send $X^i_{\mathcal{S}}=\bar{\oplus}_{s\in \mathcal{S}}W^i_{d_s,\mathcal{S}\setminus\{s\}}$}
\EndFor
\EndFor
\EndProcedure
\EndFor
\EndProcedure
\end{algorithmic}
\end{algorithm}
Using (\ref{eq:4-2}), the delivery rate of Algorithm 3 can be derived.

\begin{Theorem}
For the decentralized coded caching system described above, Algorithm 3 achieves a delivery rate given by
\begin{align}\label{eq:4-3}
R(M_1, &..., M_K)=\min\left\{\sum\limits_{i=1}^K (r_{i}-r_{i-1})\right.\sum\limits_{l=1}^{K-i+1}\prod\limits_{k=1}^l(1-t^i_k), \nonumber\\
&~~\left. \sum\limits_{i=1}^{\min\{N, K\}} r_{K-i+1}- \min\limits_{\substack{\mathcal{S}\subset\{1, ..., K\}\\|\mathcal{S}|=\min\{N, K\}}}\sum\limits_{k \in \mathcal{S}}\frac{M_k}{N}\right\},
\end{align}
where $\{t^i_1, t^i_2,..., t^i_{K-i+1}\}$ is an ordered permutation of $\left\{\frac{M_{i}}{Nr_{i}}, \frac{M_{i+1}}{Nr_{i+1}}, ..., \frac{M_{K}}{Nr_{K}}\right\}$ such that $t^i_1\leq t^i_2 \leq \cdots \leq t^i_{K-i+1}$, $i\in\{1, ..., K\}$.
\begin{proof}
We first prove the first term in (\ref{eq:4-3}), which is provided by DELIVERY 1. We sum up the rates corresponding to all possible multicasting subsets, $\mathcal{S}\subset\{1, ..., K\}$. From (\ref{eq:4-2}), we have
\begin{subequations}
\begin{align}
R(M_1, ..., M_K)&=\sum\limits_{k=1}^K&\sum\limits_{\substack{\mathcal{S}\subset\{1, ..., K\}\\|\mathcal{S}|=k}}\sum\limits_{i=1}^{m( \mathcal{S})} (r_{i}-r_{i-1})\cdot\nonumber\\
&&p^i_{(d_{r'(\mathcal{S})}, D_{r'(\mathcal{S})}),\mathcal{S}\setminus\{r'(\mathcal{S})\}}\label{rate11}\\
&=\sum\limits_{i=1}^K& \sum\limits_{k=1}^{K-i+1}\sum\limits_{\substack{\mathcal{S}\subset\{i, ..., K\}\\|\mathcal{S}|=k}} (r_{i}-r_{i-1})\cdot\nonumber\\
&& p^i_{(d_{r'(\mathcal{S})}, D_{r'(\mathcal{S})}),\mathcal{S}\setminus\{r'(\mathcal{S})\}}, \label{rate22}
\end{align}
\end{subequations}
where $k$ denotes the cardinality of subsets, for $k=1, ..., K$. Note that $n \cdot(r_{i}-r_{i-1}) p^i_{(d_{r'(\mathcal{S})}, D_{r'(\mathcal{S})}),\mathcal{S}\setminus\{r'(\mathcal{S})\}}$ denotes the number of bits from the $i$th layer in $|X_{\mathcal{S}}|$. The above equation implies that it is equivalent to delivering each layer separately. We present DELIVERY1 as the layered content delivery approach in Algorithm 4, where $W^i_{d_s,\mathcal{S}\setminus\{s\}}$ denotes the set of bits from the $i$th layer of file $d_s$ that are cached exclusively by users in the subset $\mathcal{S}\setminus\{s\}$. Hence, 
\begin{equation}
|X^i_{\mathcal{S}}|=n\cdot (r_{i}-r_{i-1}) p^i_{(d_{r'(\mathcal{S})}, D_{r'(\mathcal{S})}),\mathcal{S}\setminus\{r'(\mathcal{S})\}}.
\end{equation}
We focus on the delivery of the $i$th layer to users $\{i, ..., K\}$ in the following. For this layer, we order and re-index users $\{i, ..., K\}$ with $\{1, .., K-i+1\}$ such that $\frac{M_k}{Nr_k}=t^i_k$, for $k=1, ..., K-i+1$. Based on this re-indexing, we have
\begin{align}\label{eq:4-4}
\sum\limits_{k=1}^{K-i+1}\sum\limits_{\substack{\mathcal{S}\subset\{i, ..., K\}\\|\mathcal{S}|=k}}& p^i_{(d_{r'(\mathcal{S})}, D_{r'(\mathcal{S})}),\mathcal{S}\setminus\{r'(\mathcal{S})\}}\nonumber\\
&=\sum\limits_{l=1}^{K-i+1}\prod\limits_{k=1}^l(1-t^i_k),
\end{align}
which yields the first term in (\ref{eq:4-3}). A detailed proof of (\ref{eq:4-4}) can be found in Appendix D. 

Now, we continue with analysis of the DELIVERY2 procedure in Algorithm 3. Note that the message sent at step 17 of Algorithm 3 is targeted at a group of users requesting file $S_i^n$, for $i=1, ..., N$. Let $\mathcal{S}_i$ denote the set of users requesting file $S_i^n$, $i=1, ..., N$. Since each user $k \in \mathcal{S}_i$ has $nM_k/N$ bits of file $S_i^n$ stored in its caches, with the fact stated in \cite[Appendix A]{maddah2013decentralized}, at most 
\begin{equation}
n\left\{\max\limits_{k \in \mathcal{S}_i}r_k-\min\limits_{k \in \mathcal{S}_i}\frac{M_k}{N}\right\}
\end{equation}
random linear combinations need to be sent for all those users in $\mathcal{S}_i$ to decode file $S_i^n$ at their desired distortion values. 
If $N\geq K$, the worst case demand combination corresponds to each user requesting a distinct file. There are at most $K$ files requested, which yields a delivery rate of 
\begin{equation}\label{ngkrandom}
\sum\limits_{i=1}^{K} \left\{r_{i}- \frac{M_i}{N}\right\}
\end{equation}
for the DELIVERY2 procedure.

If $N< K$, the worst case demand combination occurs when the $N$ users with the lowest distortion requirements, i.e., the $N$ largest $r_k$ values, $k=1, ..., K$, and the $N$ users with the smallest cache capacities are in different groups $\mathcal{S}_i$, $i=1, ..., N$, that is, they request $N$ distinct files. Since $D_1 \geq D_2 \geq \cdots \geq D_k$, i.e., $r_1\leq r_2\leq \cdots \leq r_K$, for this case, the DELIVERY2 procedure achieves a delivery rate of 
\begin{equation}\label{nlkrandom}
\sum\limits_{i=1}^{N} r_{K-i+1}- \min\limits_{\substack{\mathcal{S}\subset\{1,., K\}\\|\mathcal{S}|=N}}\sum\limits_{k \in \mathcal{S}}\frac{M_k}{N}.
\end{equation}

Combing (\ref{ngkrandom}) and (\ref{nlkrandom}), we have the second term in (\ref{eq:4-3}). This completes the proof.
\end{proof}
\end{Theorem}

We remark that Eqn. (\ref{eq:4-4}) denotes the probability of any bit of the $i$th layer to be sent in the delivery phase. Note that (\ref{eq:4-4}) has a similar form to the expression of the achievable delivery rate in the lossless coded caching problem with heterogeneous cache sizes presented in \cite[Theorem 3]{wang2015heterogeneouscachesizes}. This implies that, with the \textit{ placement phase} carried out as in Algorithm 3, the delivery rate is equivalent to the lossless coded caching scheme proposed in \cite{wang2015heterogeneouscachesizes} for each layer.

\begin{algorithm}\label{alg5}
\caption{Layered Content Delivery 2 (LCD2)(Based on Algorithm 1 in \cite{Amiri2016heterogeneouscachesizes})}
\begin{algorithmic}[1]
\Statex
\Procedure {DELIVER $(d_1, ..., d_K)$}{}
\For {$i=1, ..., K$}
\Procedure {delivery $i$th layer of files $(d_i, ..., d_K)$ to users $\{i, ..., K\}$}{}
\State{Define $K_j$ as the number of users in $\{i, ..., K\}$ that requests file $j$, for $j=1, ..., N$;}
\State{Reorder and reindex users $\{i, ..., K\}$ such that $d_k=j$, for $j=1, ..., N$, and $k=S_{j-1}+1, ..., S_j$ where $S_0=i-1$ and $S_j=\sum\limits_{l=1}^j K_l$;}
\State{\textbf{Part 1}: Delivering bits that are not in the cache of any user}
\For {$j = 1, 2, \ldots, N$}
\State{$X^i_1=\left( W^i_{d_{S_{j-1} + 1},\left\{ \emptyset \right\}}  \right)$}
\EndFor
\Statex
\State{\textbf{Part 2}: Delivering bits that are in the cache of only one user}
\State{\begin{align}X^i_{21}=\left( {\bigcup\limits_{j = 1}^N {\bigcup\limits_{k = {S_{j - 1}} + 1}^{{S_j} - 1} {\left( {{W^i_{j,\left\{ k \right\}}} \bar \oplus {W^i_{j,\left\{ k + 1 \right\}}}} \right)} } } \right)\nonumber\end{align}}
\State{\begin{align}~~~~~X^i_{22}=\bigcup\limits_{j = 1}^{N - 1}& \bigcup\limits_{h = j+ 1}^N \left( \bigcup\limits_{k = {S_{h - 1}} + 1}^{{S_h} - 1} {\left( {{W^i_{j,\left\{ k \right\}}} \bar \oplus {W^i_{j,\left\{ {k + 1} \right\}}}} \right)},\right.\nonumber\\&\left.\bigcup\limits_{k = {S_{j - 1}} + 1}^{{S_j} - 1} {\left( {{W^i_{h,\left\{ k \right\}}} \bar \oplus {W^i_{h,\left\{ {k + 1} \right\}}}} \right)} \right)\nonumber\end{align}}

\State{\begin{align}X^i_{23}=\left( \bigcup\limits_{j = 1}^{N - 1} \bigcup\limits_{h = j + 1}^N W^i_{j,\left\{S_{h-1}+1\right\}} \bar \oplus W^i_{h,\left\{S_{j-1}+1\right\}}   \right)\nonumber\end{align}}
\Statex
\State{\textbf{Part 3}: Delivering bits that are in the cache of more than one user}
\For{$j = i,  \ldots, K - 2$ }
\For{$h = 2, 3, \ldots, K - i$ }
\For{$V \subset \left[ {j + 1:K} \right]: \left| V \right| = h$ }
\State{\begin{align}{X^i_3} = \left( {\left( {\mathop {\bar  \oplus }\limits_{v \in V} {W^i_{{d_v},\left\{ {V,j} \right\}\backslash \left\{ v \right\}}}} \right)\bar  \oplus {W^i_{{d_j},V}}} \right)\nonumber\end{align}}
\EndFor
\EndFor
\EndFor
\EndProcedure
\EndFor
\EndProcedure
\Statex
\end{algorithmic}
\end{algorithm}

For the decentralized lossless coded caching problem with distinct cache capacities, \cite{Amiri2016heterogeneouscachesizes} further exploits the multicasting gain among users requesting the same file, which achieves a lower delivery rate when the number of users is larger than the number of files, compared to \cite{wang2015heterogeneouscachesizes}.  Here, in Algorithm 5, we employ a caching strategy similar to \cite[Algorithm 1]{Amiri2016heterogeneouscachesizes}. Based on \cite[Theorem 1]{Amiri2016heterogeneouscachesizes}, we have Theorem 5 specifying the delivery rate achieved by Algorithm 5.

\begin{Theorem}
For the decentralized coded caching system described above, Algorithm 5 achieves a delivery rate given by
\begin{align}\label{eq:4-6}
&R(M_1, ..., M_K)\nonumber\\
&=\sum\limits_{i=1}^K (r_{i}-r_{i-1})\left(\sum\limits_{l=1}^{K-i+1}\prod\limits_{k=1}^l(1-t^i_k)-\Delta {R_i^1}-\Delta {R_i^2}\right),
\end{align}
where $t_k^i$ is defined as in Theorem 1, for $k\in \{1, ..., K-i+1\}$ and $i \in \{1, ..., K\}$, and

\begin{subequations}
\label{DeltaR}
\begin{align}\label{DeltaRone}
&\Delta {R_i^1}=\begin{cases}
 \left( {L_i-N} \right)\prod\limits_{l = 1}^{L_i} {\left( {1 - t_l^i}\right)},~~~~~~~~~~~&\mathrm{if}~~~~L_i>N\\
0,~~~~~~~~~~~~&\mathrm{if}~~~~L_i\leq N
\end{cases}\\
&\Delta {R_i^2} =\begin{cases}
\left[ {\sum\limits_{k = 1}^{L_i- N} {\left( {\frac{ (k-1) t_{k+N}^i}{{1 - t_{k+N}^i}}} \right)} } \right]\prod\limits_{l = 1}^{L_i} {\left( {1 - t_l^i}\right)},&\mathrm{if}~~~~L_i>N\\
0,&\mathrm{if}~~~~L_i\leq N
\end{cases}
\label{DeltaRtwo}
\end{align}
\end{subequations}
and $L_k=K-i+1$.
\end{Theorem}

We can see that if $K\leq N$, $L_i\leq N$ holds for $i=1, ..., K$, thus (\ref{eq:4-6}) is equivalent to (\ref{eq:4-3}), that is, LCD2 has the same performance with LCD1. For the case $K>N$, we have $L_i> N$ for $i=1, ..., K-N$, and LCD2 results in a reduction in the delivery rate for these layers, quantified by $\Delta {R_i^1}$ and $\Delta {R_i^2}$, in (\ref{DeltaRone}) and (\ref{DeltaRtwo}), respectively. Numerical comparison of the performance of these two content delivery algorithms, LCD1 and LCD2, is presented in Section~\ref{sec5.2}.

\section{Numerical results}\label{sec5.2}
In this section, we numerically compare the delivery rates of the proposed centralized and decentralized coded lossy caching schemes. We particularly consider two different cases depending on the relative numbers of users and files in the system, i.e., $N\geq K$ and $N<K$, as the proposed caching schemes exhibit different behaviors in these cases. 

In the first scenario, we consider a caching system with a server containing $N=10$ files serving  $K=10$ users. The target distortion levels of the users are given by $(D_1, D_2, ..., D_{10})$, which we assume to be achievable by a scalable code with rates $(r_1, r_2, ..., r_{10})= (1, 2, ..., 10)$. To compare the achievable rates with the lower bound, we assume the underlying source distribution is successively refinable, that is, $r_k=r^*_k$, $k=1, ..., K$. We consider two cases for the cache sizes: the first one with identical cache capacities, i.e.,  $M_1=M_2=\cdots=M_{10}=M$, and the second one with heterogeneous cache capacities, where $M_k=0.2kM$, for $k=1, ..., 10$. The results for these two scenarios for both the centralized and decentralized caching are plotted in Fig. 4 and Fig. 5, respectively.

\begin{figure}[t]
\centering
\includegraphics[width=1.05\linewidth]{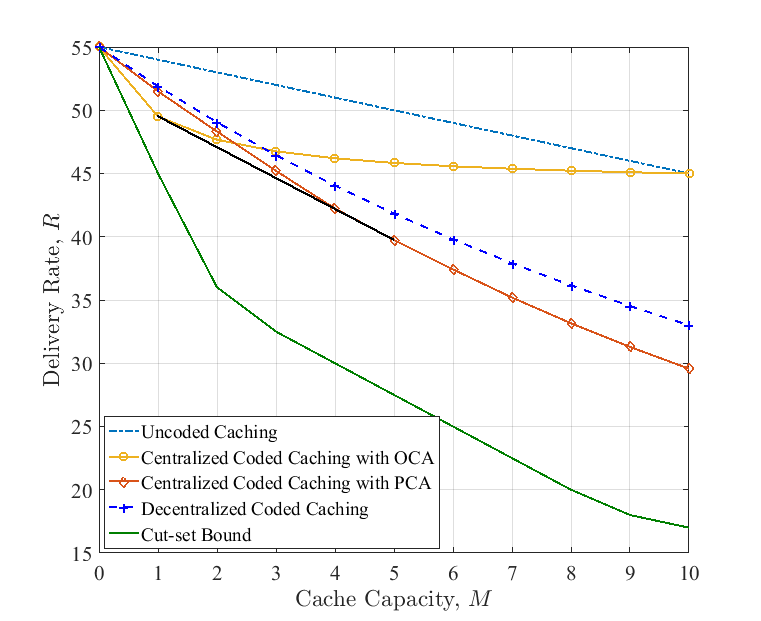}
\caption{Comparison of the achievable delivery rates with identical cache capacities, i.e., $M_1=M_2=\ldots=M_K$, for $N=10$, $K=10$.}
\end{figure}

\begin{figure}[t]
\centering
\includegraphics[width=1.018\linewidth]{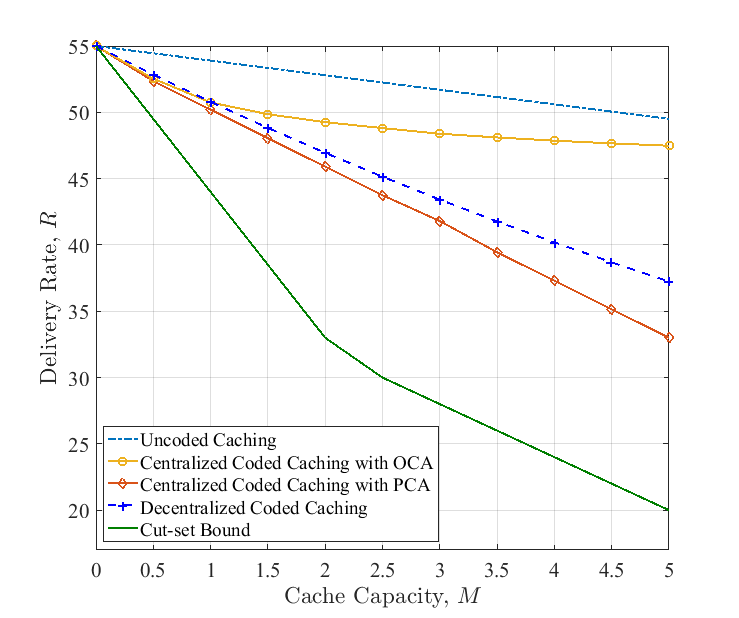}
\caption{Comparison of the achievable delivery rates with heterogeneous cache capacities, i.e., $M_k=0.2kM$, $k=1, ..., K$, for $N=10$, $K=10$.}
\end{figure}
In Fig. 4, we observe that the centralized coded caching scheme with OCA achieves the best delivery rate when the cache capacities are very small but its performance approaches that of the uncoded caching scheme as $M$ increases. This implies that when the cache capacity is small, it should be allocated mainly to the first layer, which is requested by all the users. Since the cache capacities are identical, coded caching of the same layer across users creates more multicasting opportunities, which better exploits the limited cache capacity. On the other hand, for medium to large cache capacities, PCA significantly outperforms OCA. This is because, when there is sufficient cache capacity, caching higher layers creates new multicasting opportunities, which further reduces the delivery rate. The black line in the figure is achievable by memory sharing between the two caching schemes, OCA and PCA, further reducing the delivery rate for moderate cache capacities. In Fig. 5, a network with heterogeneous cache capacities is considered, and it is observed that PCA outperforms OCA for all values of $M$, since PCA is capable of exploiting the additional cache capacity of users to meet the requirements of reduced distortion target, retaining the symmetry among the amount of cache allocated to each layer across different users. 

We remark that, since $N=K$, decentralized caching with LCD1 and LCD2 have the same performance as characterised in Theorem 4. For very small cache capacities, the decentralized scheme achieves almost the same performance as the centralized scheme with PCA. This is because, when cache capacities are very small, users will cache distinct bits from their required layers with high probability despite random placement, which is similar to the cache placement phase used by the centralized scheme with PCA. The performance improvement of centralized caching becomes more pronounced as $M$ increases since the collaboration between users during the placement phase will fully exploit the cache capacities to create the maximum number of multicasting opportunities. We observe that, despite lack of cache coordination across users, decentralized caching still achieves a performance not far from the best centralized caching scheme.

In both Fig. 4 and Fig. 5, it is observed that a significant improvement can be achieved by coded caching. We also remark that the total cache capacity across the whole network is $10M$ and $11M$ for the settings considered in Fig. 4 and Fig. 5, respectively. However, we can notice that the delivery rate achieved in Fig. 4 is significantly higher in both the centralized and decentralized scenarios. This is due to the distribution of the cache capacity across the users. In the latter scenario, the users with lower distortion requirements have larger cache capacities, allowing them to achieve their desired QoS targets without increasing the delivery rate. 
\begin{figure}[t]
\centering
\includegraphics[width=0.98\linewidth]{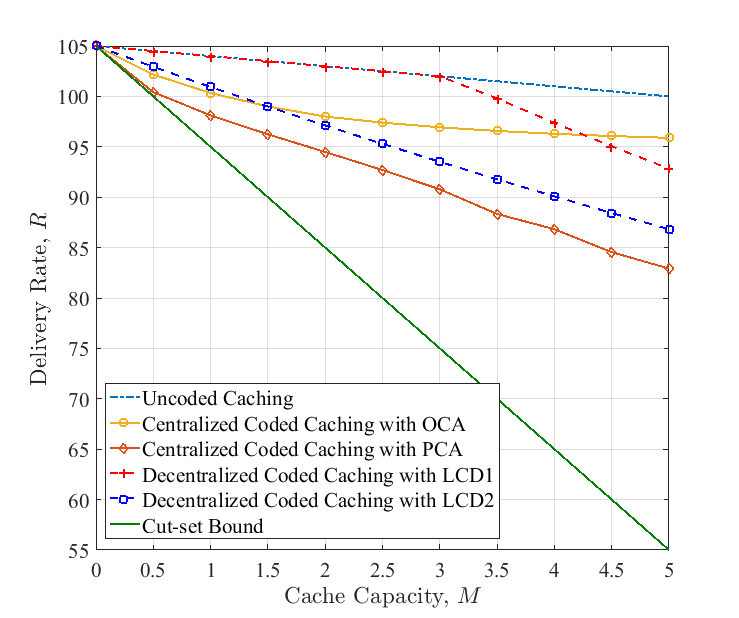}
\caption{Comparison of the achievable delivery rates with identical cache capacities, i.e., $M_1=M_2=\ldots=M_K$, for $N=10$, $K=15$.}
\end{figure}

\begin{figure}[t]
\centering
\includegraphics[width=0.98\linewidth]{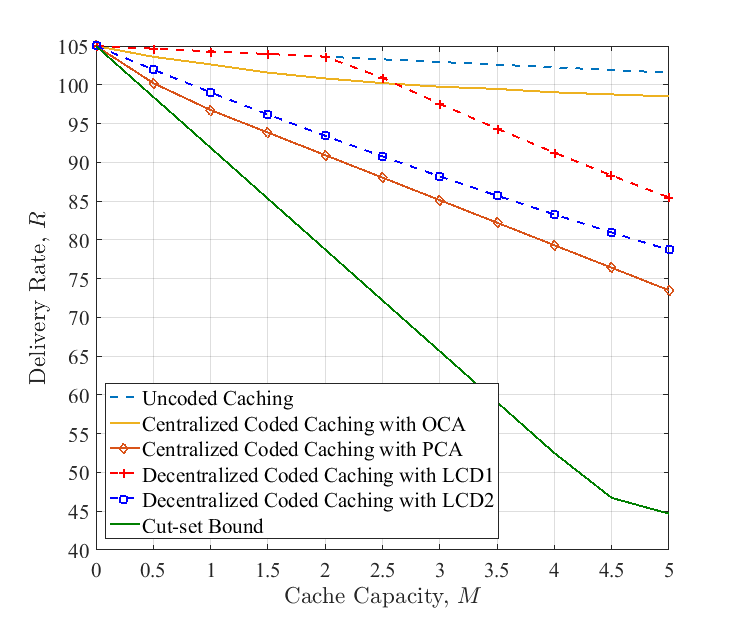}
\caption{Comparison of the achievable delivery rates with heterogeneous cache capacities, i.e., $M_k=0.125kM$, $k=1, ..., K$, for $N=10$, $K=15$.}
\end{figure}

In the second scenario, we consider a server with $N=10$ files serving $K=15$ users. The target distortion levels are such that $(r_1, r_2, ..., r_{15})= (1, 2, ..., 15)$. As in the first scenario, we consider two cases: identical cache sizes, i.e.,  $M_1=M_2=\cdots=M_{15}=M$, plotted in Fig. 6; and heterogeneous cache sizes, i.e., $M_k=0.125kM$, for $k=1, ..., 15$, plotted in Fig. 7, such that the total cache capacity across the network for both scenarios is $15M$.

We observe that, for centralized coded caching, PCA outperforms OCA in both Fig. 6 and Fig. 7. This implies that it is always better for each user to distribute its cache capacity to all the layers it may require, rather than to allocate it only to the first layer. Since $N<K$, the number of users requesting the first layer is larger than the number of files. There are at least two users requesting the same file, which creates multicasting opportunities even with uncoded caching, reducing the gain of coded caching over uncoded caching. Allocating the cache capacities to other layers will better exploit coded caching. In decentralized caching, we see that, LCD1 has the same performance as uncoded caching for small cache capacities. However, Fig. 6 shows a larger range of cache capacities where LCD1 and uncoded caching have the same performance, since when the cache capacities are identical, for each layer, the expected number of bits cached by each user is distinct, which reduces the gain from coded delivery. It is observed in both figures, LCD2 greatly outperforms LCD1, but the gap between two schemes reduces with the cache capacity. This is because, the improvement of LCD2 over LCD1 derives from more effectively delivering the bits that are cached by at most one user. When the cache capacity increases, the number of such bits reduces, hence, the performance of LCD2 approaches that of LCD1. Although the total cache capacity across the network is the same for the scenarios considered in Fig. 6 and Fig. 7, it is observed that in both the centralized and decentralized settings, the rates achieved in Fig. 7 are significantly lower, similar to the $N=K$ scenario, since the larger cache capacities can be exploited to improve the QoS of users by the proposed layered caching approach.

We also observe that in all the figures, the gain of coded caching, both in centralized and decentralized settings, becomes more significant as the cache capacity, $M$, increases. We also note that there is a relatively large gap between the best achievable delivery rates and the cut-set lower bound, but part of this gap is potentially due to the looseness of the cut-set bound, as also suggested in \cite{maddah2014fundamental}.

\begin{figure}[t]
\centering
\includegraphics[width=1.08\linewidth]{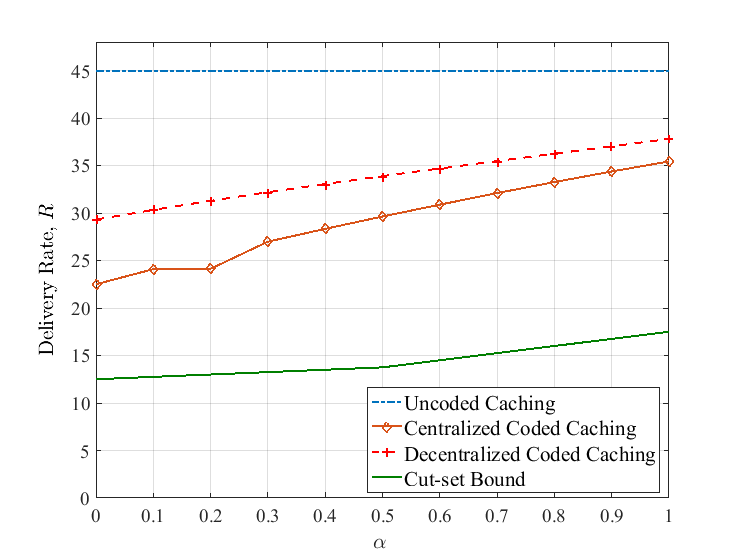}
\caption{Comparison of achievable delivery rates versus $\alpha$, $\alpha \in [0, 1]$, $M_1=M_2=\cdots=M_{10}=5$, for $N=10$, $K=10$.}
\end{figure}

Next, we consider a server with $N=10$ files serving $K=10$ users, with identical cache capacities, i.e., $M_1=M_2=\cdots=M_{10}=5$. The target distortion levels of the users, $(D_1, D_2, \ldots, D_{10})$, are given by $r_k=5+(k-5.5)\alpha$, for $k=1, ..., 10$ and $\alpha \in [0, 1]$. In this scenario, the average value of $r_k$ is $5$, i.e., $\sum\limits_{k=1}^{10} r_k/10=5$, independent of the value of $\alpha$. As $\alpha$ increases, the distortion requirements of users become more diverse. The delivery rates of the proposed caching schemes, in centralized and decentralized setting, are shown in Fig. 8, compared with \textit{uncoded caching} and cut-set bound. We emphasize that the delivery rate of the centralized caching scheme in Fig. 8 is the lower one of the delivery rates achieved by PCA and OCA. We observe that, while the delivery rate of \textit{uncoded caching} remain the same, the delivery rates of both centralized and decentralized coded caching schemes increase with $\alpha$, which indicates the loss of coded caching gain due to the diversity of distortion requirements. 
\begin{figure}[t]
\centering
\includegraphics[width=1.09\linewidth]{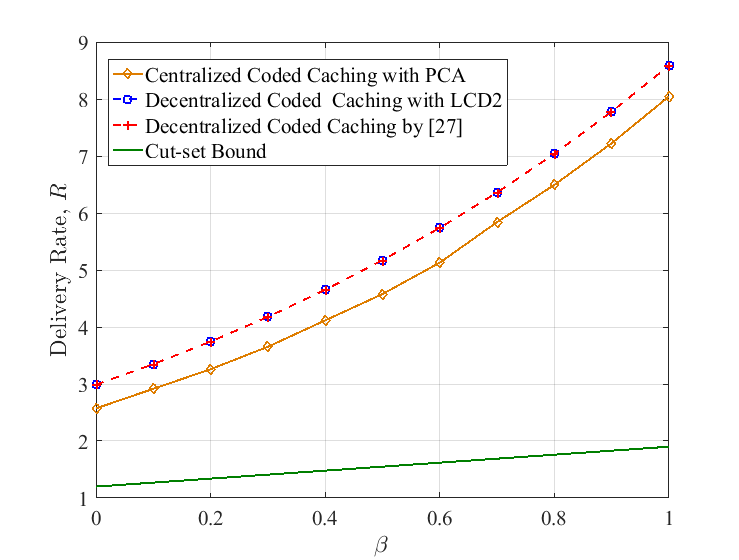}
\caption{The achievable delivery rates versus $\beta$, $\beta \in [0, 1]$, $r_1=r_2=\cdots=r_{15}$, for $N=10$, $K=15$.}
\end{figure}

Finally, we consider a server with $N=10$ files, and $K=15$ users with identical QoS requirements, i.e., $D_1=D_2=\cdots=D_{15}$, such that $r_1=r_2=\cdots=r_{15}=2$. The cache size of each user is given by $M_k=8+(k-8)\beta$, for $k=1, ..., 15$ and $\beta \in [0, 1]$. The larger $\beta$, the more skewed the cache size distribution is. This setting is equivalent to a lossless coded caching problem with distinct cache capacities. The state-of-art decentralized coded caching scheme for this setting is presented in \cite{Amiri2016heterogeneouscachesizes}. In Fig. 9, we compare the best achievable rates by the proposed centralized and decentralized coded caching schemes with the one proposed in \cite{Amiri2016heterogeneouscachesizes}. Our decentralized coded caching scheme is shown to achieve the same performance as the scheme proposed in \cite{Amiri2016heterogeneouscachesizes}. This is because we adopt the scheme proposed in \cite{Amiri2016heterogeneouscachesizes} for each layer. Since there is only one layer in this scenario, it is equivalent to applying the scheme proposed in \cite{Amiri2016heterogeneouscachesizes}  to this layer. For all the three schemes, the gain from coded caching becomes more pronounced as $\beta$ becomes small, i.e., as the cache capacities of users become more similar.  

\section{Conclusions}\label{sec5}

We have considered coded caching and delivery of contents in wireless networks, taking into account the heterogeneous distortion requirements of users, in both centralized and decentralized settings. The caching and delivery schemes considered here exploit the specific properties of lossy reconstruction of contents by users. In particular, we have exploited scalable coding, which allowed us to cache contents incrementally across users depending on their cache capacities and distortion requirements. In the \textit{centralized} setting, for a simple setting of two users and two files, we have derived the optimal coded caching scheme that achieves the information-theoretic lower bound when the underlying source distribution is successively refinable. We have further explored the case with two users and an arbitrary number of files. The proposed scheme is proven to be optimal for successively refinable sources when the cache capacities of the two users are the same and the number of files in the database is divisible by $3$. Then, assuming that the server employs scalable coding to compress all the files in the database at the required distortion levels, we tackled the general case with $K$ users and $N$ files in two steps: delivery rate minimization, which finds the minimum delivery rate for each layer separately, and cache capacity allocation among layers. We proposed two algorithms for the latter, namely, PCA and OCA. 

In the decentralized setting, since the number and identity of the active users are not known during the \textit{placement phase}, we have employed random cache placement. We have proved that a layered delivery scheme, which delivers each layer separately, is without loss of optimality, compared to joint delivery across layers. We have applied the existing coded caching scheme for the lossless caching problem with heterogeneous cache capacities to the delivery of each layer, together with the analysis on the achievable delivery rate. We have validated the improvement of the proposed coded caching schemes compared to uncoded caching through numerical results. However, there is still a remarkable gap between the best achieved delivery rate and the cut-set lower bound in all the scenarios considered in Section V. We have derived a tighter bound for a special scenario with two users, which was then used to prove the optimality of the proposed coded caching scheme for this scenario. Extending the tighter lower bounds for the lossless caching problem proposed in \cite{SenguptaCaching, GhasemiCachingLowerBound} to the lossy caching problem studied here is currently under consideration to better understand the optimal performance.

\appendices
\section{Proof of Theorem 1}
Let $s\in \{1, ..., \min\{N, K\}\}$, and consider a set of users $\mathcal{U}$ with $|\mathcal{U}|=s$. There exists a demand combination and a corresponding message over the shared link, say $X^n_1$, such that $X^n_1$ and $\{Z^n_k\mid k\in \mathcal{U}\}$ allow the reconstruction of files $S^n_1, ..., S^n_s$, each at the required distortion level from $\{D_k\mid k\in \mathcal{U}\}$. Similarly, there exists an input to the shared link, say $X^n_2$, such that $X^n_2$ and $\{Z^n_k\mid k\in \mathcal{U}\}$ allow the reconstruction of files $S^n_{s+1}, ..., S^n_{2s}$, each at the required distortion level from $\{D_k\mid k\in \mathcal{U}\}$, and so on so forth. We can continue in the same manner considering messages $X^n_3$, ..., $X^n_{\lfloor N/s\rfloor}$ for further demand combinations. Hence, with $X^n_1$, ..., $X^n_{\lfloor N/s\rfloor}$ and $\{Z^n_k\mid k\in \mathcal{U}\}$, each user $k \in \mathcal{U}$ should be able to reconstruct a distinct set of $\lfloor N/s\rfloor$ files at the corresponding distortion level $D_k$. By considering a cut separating $X^n_1$, ..., $X^n_{\lfloor N/s\rfloor}$ and $\{Z^n_k\mid k\in \mathcal{U}\}$ from the corresponding users, we have
\begin{align}\label{theo1:proof1}
 \sum\limits_{i=1}^{\lfloor N/s\rfloor} |X^n_i|+\sum\limits_{k\in \mathcal{U}} |Z_k| \geq \lfloor N/s\rfloor \sum\limits_{k\in \mathcal{U}} R(D_k),
\end{align}
according to the cut-set bound\cite[Theorem 14.10.1]{cover2012elements}. From the definition of $R^{*}(M_1,..., M_K)$, we have $R^{*}(M_1,..., M_K) \geq |X^n_i|$, for $i=1, ..., \lfloor N/s\rfloor$. We also have the cache capacity constraints $M_k \geq |Z_k|$, for $k \in \mathcal{U}$. Plugging these into \eqref{theo1:proof1}, we obtain
\begin{align}
\lfloor N/s\rfloor R^{*}(M_1,..., M_K)+\sum\limits_{k\in \mathcal{U}} M_k\geq \lfloor N/s\rfloor \sum\limits_{k\in \mathcal{U}} R(D_k).
\end{align}
Since this inequality must hold for all possible choices of $s$, and corresponding subset of users $\mathcal{U}$ with $|\mathcal{U}|=s$, we can obtain the folowing cut-set bound by substituting $r^*_k=R(D_k)$:
\begin{align}
&R^{*}(M_1,..., M_K) \geq R_{C}(M_1,..., M_K)\nonumber\\
&\triangleq \operatorname*{max}\limits_{s\in \{1,...,\min\{N, K\}\}} \operatorname*{max}\limits_{\mathcal{U}\subset \{1,...,K\}, |\mathcal{U}|=s}\left(\sum\limits_{k\in \mathcal{U}} r^*_{k}-\frac{\sum\limits_{k\in \mathcal{U}}M_k}{\lfloor N/s\rfloor}\right). 
\end{align}

\section{Proof of Theorem 2}
We consider two groups of demands, i.e., $\{(2i+1,2i+2)| i=0, 1, ..., \lfloor N/2 \rfloor-1\}$ and $\{(2i+2,2i+1)| i=0, 1, ..., \lfloor N/2 \rfloor-1\}$. We define the vector of channel inputs corresponding to these demand combinations as follows $\bar{X}^n_{12}=\bigcup\limits_{i=0}^{\lfloor N/2 \rfloor-1} X^n_{(2i+1,2i+2)}$, and $\bar{X}^n_{21}=\bigcup\limits_{i=0}^{\lfloor N/2 \rfloor-1} X^n_{(2i+2,2i+1)}$. Then, for any $(n, M_1, M_2, R)$ caching code that achieves the distortion tuple $(D_1, D_2)$, we have
\begin{subequations}
\begin{align}
2\lfloor N/2 \rfloor nR+&nM_1+nM_2\nonumber\\
             &\geq H(\bar{X}^n_{12}, Z^n_1)+ H(\bar{X}^n_{21}, Z^n_2)\\
             &= H(\bar{X}^n_{12}, Z^n_1|\bar{S}^n_1)+ H(\bar{X}^n_{21}, Z^n_2|\bar{S}^n_1)\nonumber\\
             &~~+ I(\bar{S}^n_1; \bar{X}^n_{12}, Z^n_1)+ I(\bar{S}^n_1; \bar{X}^n_{21}, Z^n_2)\\
             &\geq H(\bar{X}^n_{12}, Z^n_1, \bar{X}^n_{21}, Z^n_2|\bar{S}^n_1)\nonumber\\
             &~~+ I(\bar{S}^n_1; \bar{X}^n_{12}, Z^n_1)+ I(\bar{S}^n_1; \bar{X}^n_{21}, Z^n_2)\\
             &\geq I(\bar{S}^n_2; \bar{X}^n_{12}, Z^n_1, \bar{X}^n_{21}, Z^n_2|\bar{S}^n_1)\nonumber\\
             &~~+ I(\bar{S}^n_1; \bar{X}^n_{12}, Z^n_1)+ I(\bar{S}^n_1; \bar{X}^n_{21}, Z^n_2) \label{eq:9}
\end{align}
\end{subequations}
where $\bar{S}^n_1=\bigcup\limits_{i=0}^{\lfloor N/2 \rfloor-1} S^n_{2i+1}$ and  $\bar{S}^n_2=\bigcup\limits_{i=0}^{\lfloor N/2 \rfloor-1} S^n_{2i+2}$. We denote by $\hat{S}^n_{i,k}$ the reconstruction of $S^n_i$ at user $k$, for $i=1, ..., N$ and $k=1, 2$. Let $\hat{\bar{S}}^n_{11}=\bigcup\limits_{i=0}^{\lfloor N/2 \rfloor-1} \hat{S}^n_{2i+1,1}$,  $\hat{\bar{S}}^n_{12}=\bigcup\limits_{i=0}^{\lfloor N/2 \rfloor-1} \hat{S}^n_{2i+1,2}$ and $\hat{\bar{S}}^n_{22}=\bigcup\limits_{i=0}^{\lfloor N/2 \rfloor-1} \hat{S}^n_{2i+2,2}$. We have
\begin{subequations}\label{rd1}
\begin{align}
I(\bar{S}^n_1; \bar{X}^n_{12}, Z^n_1)&{\geq} I(\bar{S}^n_1; \hat{\bar{S}}^n_{11})\label{labela}\\
             &{\geq}\sum\limits_{i=0}^{\lfloor N/2 \rfloor-1}I(S^n_{2i+1}; \hat{\bar{S}}^n_{11})\label{labelb}\\
             &{\geq}\sum\limits_{i=0}^{\lfloor N/2 \rfloor-1}I(S^n_{2i+1}; \hat{S}^n_{2i+1,1})\\
             &{\geq}\lfloor N/2 \rfloor nR(D_1),\label{labelc}
\end{align}
\end{subequations}
where the inequality \eqref{labela} follows from the data processing inequality since $\bar{S}^n_1-(\bar{X}^n_{12}, Z^n_1)-\hat{\bar{S}}^n_{11}$ forms a Markov chain; \eqref{labelb} holds due to the independence of files. Since the reconstruction of $S^n_{2i+1}$ at user $1$, i.e., $\hat{S}^n_{2i+1,1}$, is required to be within distortion $D_1$, the inequality \eqref{labelc} follows from the definition of the rate distortion function \cite{equitz1991successive}.

Similarly, we have 
\begin{subequations}\label{rd2}
\begin{align}
I(\bar{S}^n_1; \bar{X}^n_{21}, Z^n_2)&{\geq} I(\bar{S}^n_1; \hat{\bar{S}}^n_{12})\\
&{\geq}\sum\limits_{i=0}^{\lfloor N/2 \rfloor-1}I(S^n_{2i+1}; \hat{\bar{S}}^n_{12})\\
&{\geq}\sum\limits_{i=0}^{\lfloor N/2 \rfloor-1}I(S^n_{2i+1}; \hat{S}^n_{2i+1,2})\\
&{\geq}\lfloor N/2 \rfloor nR(D_2).
\end{align}
\end{subequations}

Finally, also using the data processing inequality due to the Markov chain, $(\bar{S}^n_1, \bar{S}^n_2)-(\bar{X}^n_{12}, Z^n_2)-\hat{\bar{S}}^n_{22}$, we can write 
\begin{subequations}
\begin{align}
I(\bar{S}^n_2; \bar{X}^n_{12}, Z^n_1,\bar{X}^n_{21}, Z^n_2|\bar{S}^n_1)&{\geq} I(\bar{S}^n_2; \bar{X}^n_{12}, Z^n_2|\bar{S}^n_1)\\
&{\geq} I(\bar{S}^n_2; \hat{\bar{S}}^n_{22}|\bar{S}^n_1)\\
&{=} I(\bar{S}^n_2; \hat{\bar{S}}^n_{22},  \bar{S}^n_1)\label{prove1}\\
&{=}I(\bar{S}^n_2; \hat{\bar{S}}^n_{22}){\geq}\lfloor N/2 \rfloor nR(D_2),\label{rd3}
\end{align}
\end{subequations}
where (\ref{prove1}) holds since $\bar{S}^n_2$ is independent $\bar{S}^n_1$; and \eqref{rd3} follows from the definition of the rate-distortion function.

Substituting inequalities (\ref{rd1}), (\ref{rd2}), (\ref{rd3}) into (\ref{eq:9}) and replacing $R(D_k)$ with $r^*_k$, $k=1, 2$, we obtain
\begin{equation}
R\geq r^*_1/2+r^*_2-\frac{(M_1+M_2)}{2\lfloor N/2 \rfloor}.
\end{equation}

\section{Lower Bound of Lemma 2}

We consider two groups of $\lfloor N/3 \rfloor$ demands, i.e., $\{(3i+1,3i+2)| i=0, 1, ..., \lfloor N/3 \rfloor-1\}$ and $\{(3i+2,3i+3)| i=0, 1, ..., \lfloor N/3 \rfloor-1\}$.  Similarly to the proof of Theorem 2, we define the vector of channel inputs corresponding to these demand combination as $\bar{X}^n_{12}=\bigcup\limits_{i=0}^{\lfloor N/3 \rfloor-1} X^n_{(3i+1,3i+2)}$, and $\bar{X}^n_{23}=\bigcup\limits_{i=0}^{\lfloor N/3 \rfloor-1} X^n_{(3i+2,3i+3)}$. Then for any $(n, M_1, M_2, R)$ caching code that achieves the distortion tuple $(D_1, D_2)$, we have  
\begin{subequations}
\begin{align}
2\lfloor &N/3 \rfloor nR+nM_1+nM_2\nonumber\\
             &\geq H(\bar{X}^n_{12}, Z^n_2)+ H(\bar{X}^n_{23}, Z^n_1)\\
             &= H(\bar{X}^n_{12}, Z^n_2|\bar{S}^n_2)+ H(\bar{X}^n_{23}, Z^n_1|\bar{S}^n_2)\nonumber\\
             &~~+ I(\bar{S}^n_2; \bar{X}^n_{12}, Z^n_2)+ I(\bar{S}^n_2; \bar{X}^n_{23}, Z^n_1)\\
             &\geq H(\bar{X}^n_{12}, Z^n_2, \bar{X}^n_{23}, Z^n_1|\bar{S}^n_2)+ I(\bar{S}^n_2; \bar{X}^n_{12}, Z^n_2)\nonumber\\
             &~~+ I(\bar{S}^n_2; \bar{X}^n_{23}, Z^n_1)\\
             &\geq I(\bar{S}^n_1, \bar{S}^n_3; \bar{X}^n_{12}, Z^n_2, \bar{X}^n_{23}, Z^n_1|\bar{S}^n_2)\nonumber\\
             &~~+ I(\bar{S}^n_2; \bar{X}^n_{12}, Z^n_2)+ I(\bar{S}^n_2; \bar{X}^n_{23}, Z^n_1) \label{eq:5}
\end{align}
\end{subequations}
where $\bar{S}^n_1=\bigcup\limits_{i=0}^{\lfloor N/3 \rfloor-1} S^n_{3i+1}$, $\bar{S}^n_2=\bigcup\limits_{i=0}^{\lfloor N/3 \rfloor-1}S^n_{3i+2}$, and $\bar{S}^n_3=\bigcup\limits_{i=0}^{\lfloor N/3 \rfloor-1}S^n_{3i+3}$. We remind that $\hat{S}^n_{i,k}$ denotes the reconstruction of $S^n_i$ at user $k$, for $i=1, ..., N$ and $k=1, 2$. We define $\hat{\bar{S}}^n_{11}=\bigcup\limits_{i=0}^{\lfloor N/3 \rfloor-1} \hat{S}^n_{3i+1,1}$, $\hat{\bar{S}}^n_{21}=\bigcup\limits_{i=0}^{\lfloor N/3 \rfloor-1} \hat{S}^n_{3i+2,1}$, 
$\hat{\bar{S}}^n_{22}=\bigcup\limits_{i=0}^{\lfloor N/3 \rfloor-1} \hat{S}^n_{3i+2,2}$ and $\hat{\bar{S}}^n_{32}=\bigcup\limits_{i=0}^{\lfloor N/3 \rfloor-1} \hat{S}^n_{3i+3,2}$. Following the similar arguments as in (\ref{rd1}) and (\ref{rd2}), we have
\begin{subequations}
\begin{align}
&I(\bar{S}^n_2; \bar{X}^n_{12}, Z^n_2)\nonumber\\
             &~~ \geq I(\bar{S}^n_2; \hat{\bar{S}}^n_{22}) \geq n\lfloor N/3 \rfloor R(D_2)=n\lfloor N/3 \rfloor r^*_2, \label{eq:6}\\
&I(\bar{S}^n_2; \bar{X}^n_{23}, Z^n_1)\nonumber\\
             &~~ \geq I(\bar{S}^n_2; \hat{\bar{S}}^n_{21}) \geq n\lfloor N/3 \rfloor R(D_1)=n\lfloor N/3 \rfloor r^*_1.  \label{eq:7}
\end{align}
\end{subequations}
We also have
\begin{subequations}
\begin{align}
&I(\bar{S}^n_1, \bar{S}^n_3; \bar{X}^n_{12}, Z^n_2, \bar{X}^n_{23}, Z^n_1|\bar{S}^n_2)\nonumber\\
             & {\geq} I(\bar{S}^n_1, \bar{S}^n_3; \bar{X}^n_{12}, Z^n_2, \bar{X}^n_{23}, Z^n_1)-I(\bar{S}^n_1, \bar{S}^n_3; \bar{S}^n_2)\\
&{\geq}I(\bar{S}^n_1; \bar{X}^n_{12}, Z^n_2, \bar{X}^n_{23}, Z^n_1)+I(\bar{S}^n_3; \bar{X}^n_{12}, Z^n_2, \bar{X}^n_{23}, Z^n_1|\bar{S}^n_1)\label{f}\\
&{\geq} I(\bar{S}^n_1; \bar{X}^n_{12}, Z^n_1)+I(\bar{S}^n_3; \bar{X}^n_{23}, Z^n_2)\label{g}\\
&{\geq} I(\bar{S}^n_1; \hat{\bar{S}}^n_{11})+I(\bar{S}^n_3; \hat{\bar{S}}^n_{32})\label{h}\\
&{\geq} n\lfloor N/3 \rfloor R(D_1)+n\lfloor N/3 \rfloor R(D_2)\label{i}\\
&{=}n\lfloor N/3 \rfloor r^*_1+n\lfloor N/3 \rfloor r^*_2, \label{eq:8}
\end{align}
\end{subequations}
where (\ref{f}) follows since $\bar{S}^n_1$, $\bar{S}^n_2$ and $\bar{S}^n_3$ are independent; (\ref{g}) follows due to the nonnegativity of mutual information; and inequality (\ref{h}) is due to the data processing inequality and the fact that $\bar{S}^n_1-(\bar{X}^n_{12}, Z^n_1)-\hat{\bar{S}}^n_{11}$ and $\bar{S}^n_3-(\bar{X}^n_{23}, Z^n_2)-\hat{\bar{S}}^n_{32}$ are both Markov chains. Substituting inequalities (\ref{eq:6}), (\ref{eq:7}) and (\ref{eq:8}) into (\ref{eq:5}), we obtain
\begin{equation}
R\geq r^*_1+r^*_2-\frac{(M_1+M_2)}{2\lfloor N/3 \rfloor}
\end{equation}

\section{Proof of (\ref{eq:4-4})}
Here, we prove Eqn.~(\ref{eq:4-4}). We have
\begin{subequations}
\begin{align}
&\sum\limits_{k=1}^{K-i+1}\sum\limits_{\substack{\mathcal{S}\subset\{i, ..., K\}\\|\mathcal{S}|=k}} p^i_{(d_{r'(\mathcal{S})}, D_{r'(\mathcal{S})}),\mathcal{S}\setminus\{r'(\mathcal{S})\}}\nonumber\\
&{=}\sum\limits_{k=1}^{K-i+1}\sum\limits_{\substack{\mathcal{S}\subset\{1, .., K-i+1\}\\|\mathcal{S}|=k}} p^i_{(d_{m(\mathcal{S})}, D_{m(\mathcal{S})}),\mathcal{S}\setminus\{m(\mathcal{S})\}}\label{j}\\
&{=}\sum\limits_{k=1}^{K-i+1}\sum\limits_{l=1}^{K-i-k+2}\sum\limits_{\substack{\mathcal{S}\subset\{l+1, .., K+1-i\}\\|\mathcal{S}|=k-1}} p^i_{(d_l, D_l),\mathcal{S}}\label{k}\\
&{=}\sum\limits_{l=1}^{K-i+1}\sum\limits_{k=1}^{K-i-l+2}\sum\limits_{\substack{\mathcal{S}\subset\{l+1, .., K+1-i\}\\|\mathcal{S}|=k-1}} p^i_{(d_l, D_l),\mathcal{S}}\label{l}\\
&{=}\sum\limits_{l=1}^{K-i+1}\sum\limits_{k=1}^{K-i-l+2}\sum\limits_{\substack{\mathcal{S}\subset\{l+1, .., K+1-i\}\\|\mathcal{S}|=k-1}}\prod\limits_{s=1}^l(1-t^i_s)\cdot\nonumber\\
             &~~~~\prod\limits_{s\in \{l+1, .., K+1-i\}\setminus \mathcal{S}}(1-t^i_s)\prod\limits_{s\in \mathcal{S}}t^i_s\label{m}\\
&{=}\sum\limits_{l=1}^{K+1-i}\prod\limits_{s=1}^l(1-t^i_s)\cdot\left[\sum\limits_{k=1}^{K-i-l+2}\sum\limits_{\substack{\mathcal{S}\subset\{l+1, .., K+1-i\}\\|\mathcal{S}|=k-1}}\right.\nonumber
\\
             &\left.~~~~~~~\prod\limits_{s\in \{l+1, .., K+1-i\}\setminus \mathcal{S}}(1-t^i_s)\prod\limits_{s\in \mathcal{S}}t^i_s\right]\label{n},
\end{align}
\end{subequations}
where equality (\ref{j}) holds by ordering and re-indexing users \{i, ..., K\} to \{1, ..., K+1-i\} such that $\frac{M_k}{Nr_k}=t^i_k$, for $k= 1, ..., K+1-i$; equality (\ref{k}) comes from iterating the smallest index in the multicasting set from $1$ to $K-i-k+2$, which is the maximum that the value of the smallest index could be in a multicasting set with cardinality equals to $k$; equality (\ref{l}) is derived by switching the order of summations regarding to $l$ and $k$; equality (\ref{m}) is obtained by expressing $p^i_{(d_l, D_l),\mathcal{S}}$ as a function of $t^i_s$; since $\prod\limits_{s=1}^l(1-t^i_s)$ is independent of the value of $k$ and subset $\mathcal{S}$, equality (\ref{n}) holds. Since $\sum\limits_{j=1}^{K-i-l+2}\sum\limits_{\substack{\mathcal{S}\subset\{l+1, .., K+1-i\}\\|\mathcal{S}|=j-1}}\prod\limits_{s\in \{l+1, .., K+1-i\}\setminus \mathcal{S}}(1-t^i_s)\prod\limits_{s\in\mathcal{S}}t^i_s=1$, we have
\begin{align}
&\sum\limits_{j=1}^{K+1-i}\sum\limits_{\substack{\mathcal{S}\subset\{i, ..., K\}\\|\mathcal{S}|=j}} p^i_{(d_{r'(\mathcal{S})}, D_{r'( \mathcal{S})}),\mathcal{S}\setminus\{r'(\mathcal{S})\}}\nonumber\\
             &=\sum\limits_{l=1}^{K+1-i}\prod\limits_{s=1}^l(1-t^i_s)=\sum\limits_{l=1}^{K+1-i}\prod\limits_{k=1}^l(1-t^i_k),
\end{align}
which yields (\ref{eq:4-4}).
\bibliographystyle{unsrt} 
%\bibliography{scheduling}

\begin{IEEEbiography}[{\includegraphics[width=1in,height=1.25in,clip,keepaspectratio]{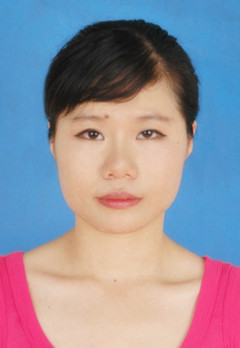}}]{Qianqian Yang}

(S'12) received the B.S. degree in automation from Chongqing University in 2011, and the M.S. degree in control engineering from Zhejiang University in 2014. She is currently a Ph.D. candidate with Imperial College London. Her current research interests include information and coding theory, resource optimization in wireless sensor networks, neural network and machine learning.

\end{IEEEbiography}

\begin{IEEEbiography}[{\includegraphics[width=1in,height=1.25in,clip,keepaspectratio]{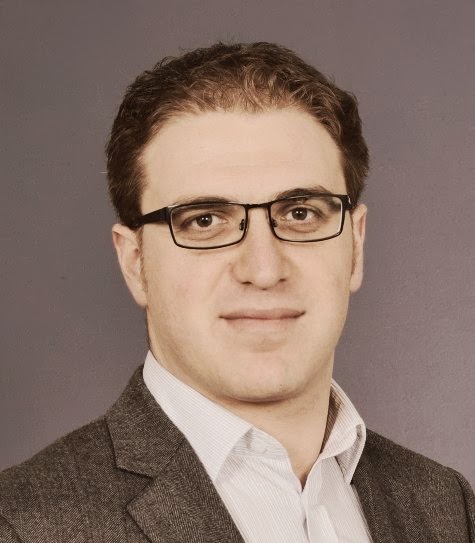}}]{Deniz G\"und\"uz}

(S'03-M'08-SM'13) received the B.S. degree in electrical and electronics engineering from METU, Turkey in 2002, and the M.S. and Ph.D. degrees in electrical engineering from NYU Polytechnic School of Engineering (formerly Polytechnic University) in 2004 and 2007, respectively. After his PhD, he served as a postdoctoral research associate at Princeton University, and as a consulting assistant professor at Stanford University. He was a research associate at CTTC in Barcelona, Spain until September 2012, when he joined the Electrical and Electronic Engineering Department of Imperial College London, UK, where he is currently a Reader in information theory and communications, and leads the Information Processing and Communications Lab.

His research interests lie in the areas of communications and information theory, machine learning, and security and privacy in cyber-physical systems. Dr. G\"und\"uz is an Editor of the IEEE TRANSACTIONS ON COMMUNICATIONS, and the IEEE TRANSACTIONS ON GREEN COMMUNICATIONS AND NETWORKING. He is the recipient of the IEEE Communications Society - Communication Theory Technical Committee (CTTC) Early Achievement Award in 2017, a Starting Grant of the European Research Council (ERC) in 2016, IEEE Communications Society Best Young Researcher Award for the Europe, Middle East, and Africa Region in 2014, Best Paper Award at the 2016 IEEE Wireless Communications and Networking Conference (WCNC), and the Best Student Paper Award at the 2007 IEEE International Symposium on Information Theory (ISIT). He is the General Co-chair of the 2018 Workshop on Smart Antennas, and previously served as the General Co-chair of the 2016 IEEE Information Theory Workshop, and a Co-chair of the PHY and Fundamentals Track of the 2017 IEEE Wireless Communications and Networking Conference. 

\end{IEEEbiography}

\end{document}